\documentclass[aip,jcp,reprint,superscriptaddress]{revtex4-1}

\usepackage{amsmath}
\usepackage{amssymb}
\usepackage{graphicx}
\usepackage{epstopdf}
\usepackage{color}
\usepackage{upgreek}
\usepackage{dcolumn}
\usepackage{bm}

\begin{document}

\title{Calculating hydrodynamic interactions for membrane-embedded objects}

 \author{Ehsan Noruzifar}
 \affiliation{Department of Chemistry and Biochemistry, University of California, Santa Barbara, Santa Barbara, USA }
 \affiliation{Department of Physics,~University of California,~Santa Barbara,~Santa Barbara,~USA}
 
 \author{Brian A. Camley}
 \affiliation{Center for Theoretical Biological Physics and Department of Physics, University of California, San Diego, La Jolla, USA}
 
 \author{Frank L.H. Brown}
 \affiliation{Department of Chemistry and Biochemistry, University of California, Santa Barbara, Santa Barbara, USA }
 \affiliation{Department of Physics,~University of California,~Santa Barbara,~Santa Barbara,~USA}

\date{\today}

\begin{abstract}
A  recently introduced numerical scheme for calculating self-diffusion coefficients of solid objects embedded in lipid
bilayer membranes is extended to enable calculation of hydrodynamic interactions between multiple
objects.  The method is used to validate recent analytical predictions by Oppenheimer and Diamant [Biophys. J., 96, 3041, 2009]
related to the coupled diffusion of membrane embedded proteins and is shown to converge to known near-field
lubrication results as objects closely approach one another, however the present methodology also applies outside of the
limiting regimes where analytical results are available.  Multiple different examples involving pairs of disk-like objects with
various constraints imposed on their relative motions demonstrate the importance of hydrodynamic interactions in the dynamics of 
proteins and lipid domains on membrane surfaces.  It is demonstrated that the relative change in self-diffusion of a
membrane embedded object upon perturbation by a similar proximal solid object displays a maximum for object sizes
comparable to the Saffman-Delbr\"uck length of the membrane.

 \end{abstract}

\maketitle

\section{Introduction}
\label{sec:intro}

Lateral motion of membrane components is required for proper physiological functioning in
cellular biology \cite{berg77,gennisbook,ram09,hell09,lingwood} and serves as fertile grounds
for biophysical studies involving lipid bilayer model systems  \cite{ott,saxton97,greenbook,keller_review,simonsvazrafts2004,brown_hydro_review,Ash2004158}.
A key aspect of dynamics taking place at the membrane surface is hydrodynamic flow, both within the membrane
itself and in the fluids surrounding the membrane.  The membrane environment is thus distinct from traditional three dimensional (3D) or
two dimensional (2D) hydrodynamic systems, incorporating aspects of both 2D and 3D flow; the membrane
components move in a two dimensional space defined by the membrane surface, but are coupled to fluid motions in a full 3D space.
For this reason, it has been suggested that membrane hydrodynamics might best be described as ``quasi-2D" \cite{Haim09}.

Saffman and Delbr\"uck \cite{saf} (SD) introduced the standard hydrodynamic model for lipid bilayer membranes.  In the SD model,
the membrane is treated as a flat thin incompressible fluid sheet with surface viscosity $\eta_m$, surrounded by an infinite incompressible aqueous bulk of viscosity $\eta$.
This model can be solved in the creeping flow limit to yield analytical predictions for the diffusion coefficient of cylindrical
bodies embedded within the membrane via no-slip boundary conditions \cite{saf,white81}, which are in good agreement
with experiments recording the diffusion of membrane proteins and lipid domains \cite{peters82,vaz82,cicuta07,petrov08,nguyen,nguyen2,krieg,ram10}.  (However, some experimental studies claim deficiencies in
the SD model for membrane protein diffusion \cite{gambin06,gambin10}.)  The
SD model also correctly predicts the dynamics of lipid domain boundary fluctuations \cite{stonemcconnell95,camleyesposito2010}
and the dynamics of phase separation in ternary model membrane systems \cite{camleybrown2010,camleybrownscaling2011,stanich2013coarsening,honerkamp2012experimental,haataja2009critical,inaura2008concentration}.

Perhaps the most concise way to express the physics contained within the SD model is through the Green's function
formulation of the hydrodynamic problem.  The membrane's velocity response to in-plane forcing
is given by 
\begin{equation}
 \label{eq:v-field}
 \mathbf{v}(\mathbf{r})= \int \mathrm{d}\mathbf{r}'\, \mathbf{T}(\mathbf{r}-\mathbf{r}') \mathbf{f}(\mathbf{r}')\,. 
\end{equation}
Here, $\mathbf{v}(\mathbf{r})$ is the $xy$ velocity of the membrane at position $\mathbf{r}=(x,y)$. (We assume the membrane
plane is coincident with the $xy$ plane of our coordinate system.)  $\mathbf{f}(\mathbf{r})$ is an in-plane  force/area applied to the 
membrane and the Green's function tensor $\mathbf{T}(\mathbf{r})$ captures the hydrodynamic response to forcing in the SD model.
Explicitly \cite{Haim09,levine02},
\begin{eqnarray}
\label{eq:mem_green}
T_{ij}(\mathbf{r}) &=& \frac{1}{4 \eta_m}\left \{ \left [ H_0(\tilde{r}) - \frac{H_1 (\tilde{r})}{\tilde{r}} -\frac{Y_0(\tilde{r})}{2} + \frac{Y_2(\tilde{r})}{2} + \frac{2}{\pi \tilde{r}^2} \right ] \right .\delta_{ij} \nonumber \\
& & \left . - \left [ H_0 (\tilde{r}) - \frac{2 H_1 (\tilde{r})}{\tilde{r}} + Y_2(\tilde{r}) + \frac{4}{\pi \tilde{r}^2} \right ] \frac{\tilde{r}_i \tilde{r}_j}{\tilde{r}^2} \right \}
\end{eqnarray}
where $Y_n$ and $H_n$ are Bessel functions of the second kind and Struve functions, respectively.
The dimensionless distance $\tilde{r} = |\mathbf{r}|/\ell_0$, where $\ell_0 = \eta_m / 2 \eta$ is
the Saffman-Delbruck length.  $\ell_0$ defines a crossover between 2D and 3D-like
hydrodynamics in the quasi-2D SD model; for separations well below $\ell_0$ the Green's function is
that of a 2D fluid with surface viscosity $\eta_m$, whereas at separations larger than $\ell_0$ the Green's function is
similar to that of a 3D system of viscosity $\eta$ \cite{lubensky96,Haim09}.  For typical lipid bilayer systems
$\ell_0$ is on the order of a micron \cite{saf}.
 
In principle, Eq. \ref{eq:v-field} is restricted to fluid regions and would not
immediately seem to be of use in solving problems involving particles suspended within the fluid (e.g. prediction of diffusion coefficients).  
However, within creeping flow, there is no physical or mathematical distinction between a solid particle embedded within the fluid via no-slip 
boundary conditions and a ``fluid" region that occupies the same physical space as the embedded particle with the supplemental restriction 
to undergo only rigid body motions.  Computational strategies that supplement Eq. \ref{eq:v-field} (typically the 3D version of it)
with the constraint of rigid-body motion over particle associated regions within the fluid have the capability to predict particulate 
dynamics.  Well known schemes that take advantage of this strategy to calculate diffusion coefficients and related properties associated
with particulate flows include the Kirkwood approximation \cite{bloomfield67,doi}, shell method
\cite{deutchbio76,swanson78} and related techniques \cite{peskin,torre,wang2013assessing}. (See Ref.~\citenum{torre} and the references within for
a review of these approaches.)  Though the majority of work in this direction has been aimed at studying various solid bodies
immersed in homogeneous 3D hydrodynamic environments, the quasi-2D membrane case has previously been considered
with a method similar to the Kirkwood approximation by Levine Liverpool and MacKintosh \cite{levine04,levine04lett}.  (It should
also be mentioned that the Green's function approach naturally leads to diffusion coefficients for circular fluid domains in the
membrane geometry \cite{dekoker,komura}.)

Though the Kirkwood approximation is simple to implement and often gives rise to qualitatively correct predictions,
the method is known to be inexact and, in extreme cases, can yield unphysical results \cite{zwanzig_weiss}.
In 3D geometries, the shell method \cite{deutchbio76,swanson78} resolves these shortcomings, however
this method is not easily applied to the SD problem as it requires use of the Rotne-Prager tensor \cite{rotne_prager};
a quasi-2D analog to the Rotne-Prager tensor has not been derived.  An appealing alternative to the shell method is
the method of Regularized Stokeslets (RS) \cite{cortez01,cortez05}. (Section \ref{sec:rs-method} provides an introduction
of the RS method.)
This technique is known to reproduce the 
correct velocity fields and particle diffusion coefficients for model problems in 3D \cite{cortez05} and is immediately extended
to the SD hydrodynamic geometry without difficulty.  In recent work ~\cite{camley13}, two of us presented this
extension and verified that RS calculations quantitatively reproduce the known analytical results 
for single-particle mobility problems (i.e. prediction of diffusion coefficients) in the quasi-2D environment.

While single-particle mobility/diffusion problems present a convenient test of computational
methodologies, the methods discussed above (RS included) are applicable to a broader
class of problems involving the motions of multiple embedded bodies and the hydrodynamic
interactions between them.  The study of many body dynamics in fluid environments has
received considerable theoretical/numerical attention \cite{montgomery77,montgomeryDeutch77,batchelor76,murphy73, 
kimbook,schmitz82,felder77,felder78,felder83,schmitz82b,felder88,mccammon,stokesian}, but
primarily in the context of 3D systems.  The few prior studies in the membrane geometry
are restricted to limiting regimes and circular particles \cite{bussel92,Haim09,henle2009effective}, due to the focus
of these works on analytical calculations.  The complicated form of Eq. \ref{eq:mem_green}
hinders analytical progress without invoking some manner of approximation.  A versatile numerical
tool to quantitatively study the dynamics of multiple membrane-embedded bodies has yet
to be developed, despite the growing experimental interest in the dynamics of solid body
suspensions in membrane systems and related quasi-2D environments
\cite{weeks06,petrov08, petrov12, knight10}.

 In this work, we extend the membrane RS method to consider many-body systems.
 A general framework is presented, which is applied in detail to the the study of 
 two circular disks within the membrane.  We also briefly revisit our earlier calculations
 \cite{camley13} involving the diffusion of diamond shaped lipid domains and tethered protein
 assemblies, as motivated by recent experiments \cite{petrov12,knight10}.

 This work is structured as follows: In Sec.~\ref{sec:dif-tensor} we discuss and 
 present the diffusion matrix for a many-body system. In Sec.~\ref{sec:rs-method} we 
 generalize the quasi-2D RS method to calculations involving multiple bodies embedded in the membrane.  This
allows for the calculation of the full many-body diffusion matrix.
 Though both sections  \ref{sec:dif-tensor} and \ref{sec:rs-method} focus primarily on the two-body case
 for concreteness, the many-body generalization is straightforward and is discussed as well. 
 In Sec.~\ref{sec:res} we present detailed results for the case of two identical disks embedded within
 the membrane. In Sections~\ref{sec:dimers} and
 \ref{sec:diamonds} we consider the case of tethered two-disk dimers and correlations between diamond shaped domains, respectively.
 Finally, Sec.~\ref{sec:summary} discusses our results and concludes.

\section{The Diffusion Matrix and Resistance Matrix}
\label{sec:dif-tensor}

The velocities and angular velocities of rigid bodies 
embedded within a fluid are related to the forces and 
torques applied to these bodies through the diffusion matrix $\mathbf{D}$ 
\cite{montgomery77,montgomeryDeutch77, felder88}: 
\begin{equation}
\label{eq:dif-tensor}
\left(
\begin{array}{cc}
\mathbf{V} \\ \boldsymbol{\Omega} 
\end{array}
\right) = \frac{\mathbf{D}}{k_B T} \left(
\begin{array}{cc}
\mathbf{F} \\ \boldsymbol{\tau} 
\end{array}
\right)
=
\frac{1}{k_BT}
\left(
\begin{array}{cc}
\mathbf{D}_\mathrm{T} & \mathbf{D}_\mathrm{TR} \\
\mathbf{D}_\mathrm{RT} & \mathbf{D}_\mathrm{R}
\end{array}
\right)
\left(
\begin{array}{cc}
\mathbf{F} \\ \boldsymbol{\tau} 
\end{array}
\right)\,.
\end{equation}
Here, $\mathbf{F}$, $\boldsymbol{\tau}$, $\mathbf{V}$ and 
$\boldsymbol{\Omega}$ contain the vector components of force, torque, velocity and angular velocity for all the bodies
present in the fluid.  $k_B$ is the Boltzmann constant ant $T$ is temperature.
This formulation, which describes an instantaneous transmission of force through an otherwise quiescent fluid, assumes an overdamped
``creeping flow" hydrodynamic regime neglecting inertial effects.  (The mobility matrix $\mathbf{M} = \mathbf{D}/k_B T$
notation is preferred by some authors; the relation between mobility and diffusion is assured by
the Einstein relation \cite{montgomery77,condiff66}.)
The blocks $\mathbf{D}_\mathrm{T}$ and $\mathbf{D}_\mathrm{R}$ in  Eq.~\eqref{eq:dif-tensor} are associated with 
translational and rotational motion, respectively. The 
blocks $\mathbf{D}_\mathrm{TR}$ and $\mathbf{D}_\mathrm{RT}$ couple
translation and rotation.  Subblocks of $\mathbf{D}_\mathrm{T}$, $\mathbf{D}_\mathrm{R}$, 
$\mathbf{D}_\mathrm{TR}$ and $\mathbf{D}_\mathrm{RT}$ associated with specific pairs of particles (e.g. $\mathbf{D}_\mathrm{T}^{1,1}$
or $\mathbf{D}_\mathrm{TR}^{3,5}$) are tensors.
Note that $\mathbf{D}_\mathrm{TR}=\mathbf{D}_\mathrm{RT}^{\intercal}$ where ``$\intercal$" denotes the transpose \cite{happel, condiff66}.

The inverse of $\mathbf{D}$ defines the resistance matrix.
\begin{equation}
\label{eq:diff-fric-tensor}
{\boldsymbol{\zeta} = k_BT \mathbf{D}^{-1}}   .
\end{equation}
The natural physical interpretation of the resistance matrix relates the {\em hydrodynamic forces/torques}
(i.e. frictions or drags) imparted to the bodies as they move through the fluid with prescribed motions
\begin{equation}
\left(
\begin{array}{cc}
\mathbf{F_{hy}} \\ \boldsymbol{\tau}_{hy} 
\end{array}
\right) = -\boldsymbol{\zeta} 
\left (
\begin{array}{cc}
\mathbf{V} \\ \boldsymbol{\Omega} 
\end{array}
\right ).
\end{equation}
However, the hydrodynamic drags exactly compensate the applied forces(torques) in creeping flow and
it is mathematically convenient to express this relationship in terms of the applied forces and
torques.  This is the formulation obtained by inverting Eq.~\eqref{eq:dif-tensor}.
\begin{equation}
\left(
\begin{array}{cc}
\mathbf{F} \\ \boldsymbol{\tau}
\end{array}
\right) =
\label{eq:fric-tensor}
\boldsymbol{\zeta}
\left (
\begin{array}{cc}
\mathbf{V} \\ \boldsymbol{\Omega} 
\end{array}
\right )
=
\left(
\begin{array}{cc}
\boldsymbol{\zeta}_\mathrm{T} &  \boldsymbol{\zeta}_\mathrm{TR} \\
\boldsymbol{\zeta}_\mathrm{RT} &  \boldsymbol{\zeta}_\mathrm{R}
\end{array}
\right)
\left (
\begin{array}{cc}
\mathbf{V} \\ \boldsymbol{\Omega} 
\end{array}
\right )
\end{equation}
with 
$\boldsymbol{\zeta}_\mathrm{TR}=\boldsymbol{\zeta}_\mathrm{RT}^\intercal$ \cite{condiff66,happel}.

Throughout this work we will be concerned with pairs of objects residing in flat membranes spanning the $xy$ plane.
Consequently, the two vectors appearing in Eqs.~\eqref{eq:dif-tensor} and ~\eqref{eq:fric-tensor} may be ordered as
$(\mathbf{F}, \boldsymbol{\tau}) \equiv (f_{1x},f_{1y},f_{2x},f_{2y},\tau_1,\tau_2)$
and $(\mathbf{V},\boldsymbol{\Omega}) \equiv (v_{1x},v_{1y},v_{2x},v_{2y},\Omega_1,\Omega_2)$ with the ${1,2}$ indices
specifying particle identity.
We will refer to these two vectors as the F-vector and V-vector, respectively.
Explicitly writing the full $6\times 6$ resistance matrix yields
(the diffusion matrix follows similarly)
\begin{equation}
  \label{eq:2-obj-fric-tensor-general}
\left(
\begin{array}{c}
 f_{1x} \\ f_{1y} \\ f_{2x} \\ f_{2y} \\ \tau_1 \\ \tau_2
\end{array}
\right)
=
\left(  
  \begin{array}{cccc|cc}
\zeta^\mathrm{11}_\mathrm{xx} & \zeta^\mathrm{11}_\mathrm{xy} & \zeta^\mathrm{12}_\mathrm{xx} & \zeta^\mathrm{12}_\mathrm{xy} & \zeta^\mathrm{11}_\mathrm{xR}&\zeta^\mathrm{12}_\mathrm{xR} \\
\zeta^\mathrm{11}_\mathrm{yx} & \zeta^\mathrm{11}_\mathrm{yy} & \zeta^\mathrm{12}_\mathrm{yx} & \zeta^\mathrm{12}_\mathrm{yy} & \zeta^\mathrm{11}_\mathrm{yR}&\zeta^\mathrm{12}_\mathrm{yR} \\
\zeta^\mathrm{21}_\mathrm{xx} & \zeta^\mathrm{21}_\mathrm{xy} & \zeta^\mathrm{22}_\mathrm{xx} & \zeta^\mathrm{22}_\mathrm{xy} & \zeta^\mathrm{21}_\mathrm{xR}&\zeta^\mathrm{22}_\mathrm{xR} \\
\zeta^\mathrm{21}_\mathrm{yx} & \zeta^\mathrm{21}_\mathrm{yy} & \zeta^\mathrm{22}_\mathrm{yx} & \zeta^\mathrm{22}_\mathrm{yy} & \zeta^\mathrm{21}_\mathrm{yR}&\zeta^\mathrm{22}_\mathrm{yR} \\
\hline
\zeta^\mathrm{11}_\mathrm{xR} & \zeta^\mathrm{11}_\mathrm{yR} & \zeta^\mathrm{12}_\mathrm{xR} & \zeta^\mathrm{12}_\mathrm{yR} & \zeta^\mathrm{11}_\mathrm{R}&\zeta^\mathrm{12}_\mathrm{R} \\
\zeta^\mathrm{21}_\mathrm{xR} & \zeta^\mathrm{21}_\mathrm{yR} & \zeta^\mathrm{22}_\mathrm{xR} & \zeta^\mathrm{22}_\mathrm{yR} & \zeta^\mathrm{21}_\mathrm{R}&\zeta^\mathrm{22}_\mathrm{R}
\end{array}
  \right)
\left(
\begin{array}{c}
 v_{1x} \\ v_{1y} \\ v_{2x} \\ v_{2y} \\ \Omega_1 \\ \Omega_2
\end{array}
\right)
\,,
\end{equation}
where the subscript $\mathrm{R}$ denotes elements associated with the 
rotational motion. 
The lines in 
Eq.~\eqref{eq:2-obj-fric-tensor-general} divides the friction tensor into
$\boldsymbol{\zeta}_\mathrm{T}$, $\boldsymbol{\zeta}_\mathrm{R}$, $\boldsymbol{\zeta}_\mathrm{TR}$ and $\boldsymbol{\zeta}_\mathrm{RT}$ 
blocks, 
see Eq.~\eqref{eq:fric-tensor}.

\begin{center}
\begin{figure}
 \includegraphics[scale=0.6]{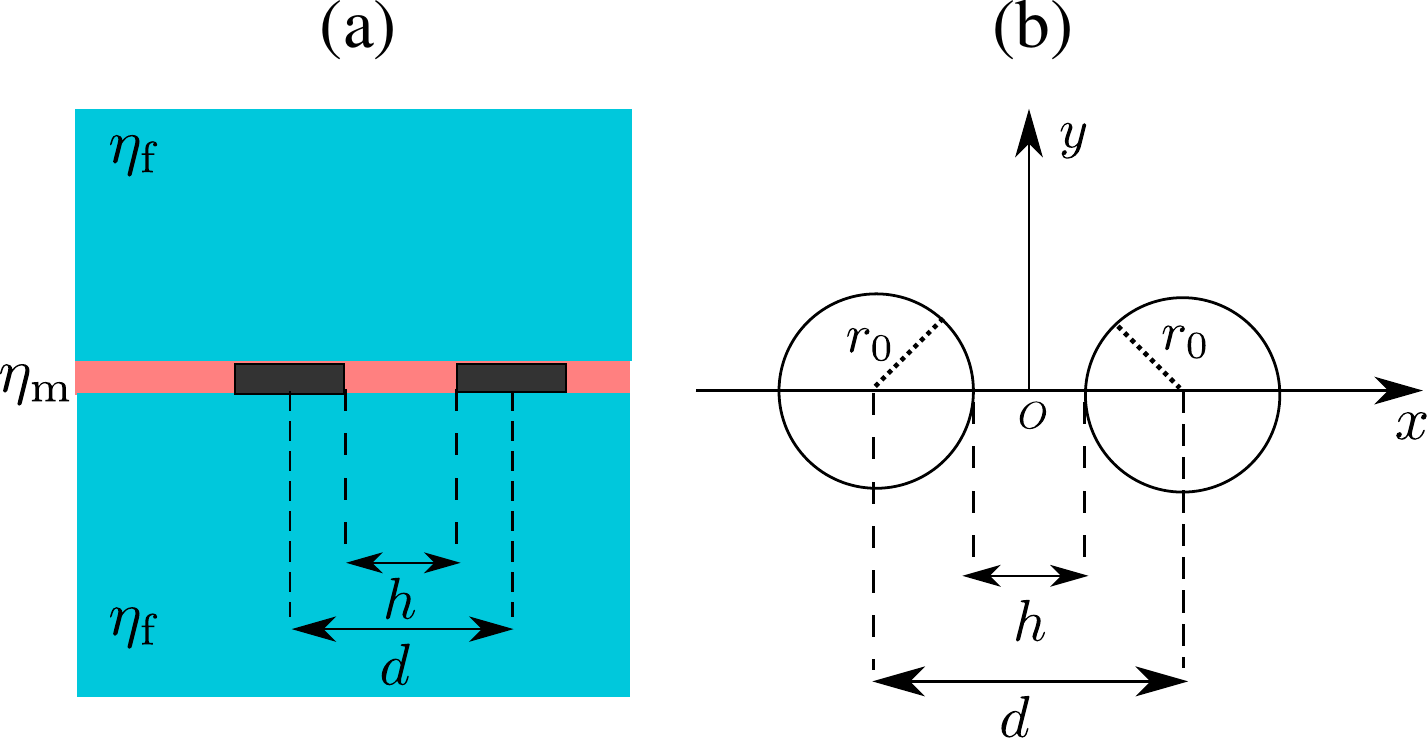}
 \caption{\label{fig:topside} 
 Schematic of the two disk system.
 (a) Side view: The two disks reside in a Saffman-Delbr\"uck membrane, which treats the bilayer as a thin structureless sheet with surface viscosity $\eta_\mathrm{m}$ 
surrounded by a bulk fluid with viscosity $\eta_\mathrm{f}$.
 (b) Top view: The disks have radius $r_0$ and center-to-center 
distance $d$; equivalently, the closest approach separation is $h=d-2r_0$. }
 \end{figure}
\end{center}

Most of the calculations in this work involve two identical disks of radius $r_0$, 
separated by a center-to-center distance $d$  (see Fig.~\ref{fig:topside}).  
This 
situation introduces significant symmetry relative to the general case and the diffusion and 
resistance matrices simplify considerably \cite{bussel92}.  Without loss of generality, we assume the
disk centers lie on the $x$ axis, which yields
\begin{equation}
  \label{eq:circ-r-mat}
\boldsymbol{\zeta}=\left(
  \begin{array}{cccc|cc}
\zeta^\mathrm{s}_\mathrm{L} & 0 & \zeta^\mathrm{c}_\mathrm{L} & 0 & 0 & 0 \\
0 & \zeta^\mathrm{s}_\mathrm{T} & 0 & \zeta^\mathrm{c}_\mathrm{T} &\zeta^\mathrm{s}_\mathrm{RT}  & \zeta^\mathrm{c}_\mathrm{RT} \\
\zeta^\mathrm{c}_\mathrm{L} & 0 & \zeta^\mathrm{s}_\mathrm{L} & 0 & 0 & 0 \\
0 & \zeta^\mathrm{c}_\mathrm{T} & 0 & \zeta^\mathrm{s}_\mathrm{T} &-\zeta^\mathrm{c}_\mathrm{RT}  & -\zeta^\mathrm{s}_\mathrm{RT} \\
\hline
0 & \zeta^\mathrm{s}_\mathrm{RT} & 0 & -\zeta^\mathrm{c}_\mathrm{RT} &\zeta^\mathrm{s}_\mathrm{R}  & \zeta^\mathrm{c}_\mathrm{R} \\
0 & \zeta^\mathrm{c}_\mathrm{RT} & 0 & -\zeta^\mathrm{s}_\mathrm{RT} &\zeta^\mathrm{c}_\mathrm{R}  & \zeta^\mathrm{s}_\mathrm{R}\\
\end{array}
  \right)\,,
\end{equation}
and
\begin{equation}
  \label{eq:circ-dif-mat}
\mathbf{D}=\left(
  \begin{array}{cccc|cc}
D^\mathrm{s}_\mathrm{L} & 0 & D^\mathrm{c}_\mathrm{L} & 0 & 0 & 0 \\
0 & D^\mathrm{s}_\mathrm{T} & 0 & D^\mathrm{c}_\mathrm{T} &D^\mathrm{s}_\mathrm{RT}  & D^\mathrm{c}_\mathrm{RT} \\
D^\mathrm{c}_\mathrm{L} & 0  &D^\mathrm{s}_\mathrm{L} & 0 & 0 & 0 \\
0 & D^\mathrm{c}_\mathrm{T} & 0 & D^\mathrm{s}_\mathrm{T} &-D^\mathrm{c}_\mathrm{RT}  & -D^\mathrm{s}_\mathrm{RT} \\
\hline
0 & D^\mathrm{s}_\mathrm{RT} & 0 & -D^\mathrm{c}_\mathrm{RT} &D^\mathrm{s}_\mathrm{R}  & D^\mathrm{c}_\mathrm{R} \\
0 & D^\mathrm{c}_\mathrm{RT} & 0 & -D^\mathrm{s}_\mathrm{RT} &D^\mathrm{c}_\mathrm{R}  & D^\mathrm{s}_\mathrm{R}\\
\end{array}
  \right)\,.
\end{equation}
Here, we have introduced the subscripts $L$ and $T$ to denote motions longitudinal or transverse to the
disk separation axis and the superscripts $s$ and $c$ to denote self and coupling components to the matrices.

In our numerical calculations for the friction and diffusion matrices (see below), we do not 
impose any of the above symmetries and calculate all 36 elements independently.
However, our results display the above symmetries to very high 
numerical precision (e.g. the ``zero" elements in Eqs. \ref{eq:circ-r-mat} and \ref{eq:circ-dif-mat} are calculated to be $10^{-10}$ to $10^{-12}$ times smaller than the 
 ``nonzero" elements in these matrices.), which provides a welcome check of our numerical algorithms.

The elements of the diffusion matrix predict the thermal fluctuations in position and orientation of the disks during a short time interval $\Delta t$
in the absence of any applied forcing.  Generically, we express this as $2D_{\alpha \beta} \Delta t = \langle\Delta q_\alpha \Delta q_\beta \rangle$ with $q_{\alpha (\beta)}$
representing any of the coordinates or angular coordinates in our system and it is understood that $\Delta t$ must be short enough to ensure that
the observed $q_{\alpha(\beta)}$ fluctuations are too small to lead to significant changes in the elements of $\mathbf{D}$.  More specifically: ($i=1,2$) 
\begin{eqnarray}
2D^\mathrm{s}_\mathrm{L}\Delta t &=& \langle (\Delta x_i)^2\rangle \nonumber \\
2D^\mathrm{c}_\mathrm{L} \Delta t &=& \langle\Delta x_1\Delta x_2 \rangle \nonumber \\ 
2D^\mathrm{s}_\mathrm{T}\Delta t &=& \langle (\Delta y_i)^2\rangle \nonumber \\
2D^\mathrm{c}_\mathrm{T}\Delta t &=& \langle\Delta y_1\Delta y_2 \rangle\nonumber \\
2D^\mathrm{s}_\mathrm{RT}\Delta t &=& \langle \Delta y_1 \Delta \theta_1\rangle = -\langle \Delta y_2 \Delta \theta_2\rangle\nonumber \\ 
2D^\mathrm{c}_\mathrm{RT}\Delta t &=& \langle\Delta y_1\Delta \theta_2 \rangle = -\langle\Delta y_2\Delta \theta_1 \rangle\nonumber \\ 
2D^\mathrm{s}_\mathrm{R}\Delta t &=& \langle (\Delta \theta_i)^2 \rangle\nonumber \\ 
2D^\mathrm{c}_\mathrm{R}\Delta t &=& \langle\Delta \theta_1\Delta \theta_2 \rangle. \label{eq:fluctuations}
\end{eqnarray}
This provides a possible experimental route toward measuring the elements of $\mathbf{D}$ that does
not involve directly perturbing the system \cite{Haim09}.

\section{Regularized Stokeslets Method for a Many-Body System}
\label{sec:rs-method}
In previous work \cite{camley13}, two of us introduced a Regularized Stokeslets \cite{cortez01,cortez05} (RS)
numerical scheme for calculating the diffusion and resistance matrices for a single
solid body embedded within a membrane.  The present paper extends this method
to calculations involving multiple solid bodies.  This section briefly reviews the 
membrane RS method and elaborates on those aspects of the calculation necessary
to consider multiple bodies.  For details on the numerical implementation, the reader
is referred to our original treatment \cite{camley13}.  

 The starting point of the RS
method is a discretized version of Eq. \ref{eq:v-field}
 \begin{equation}
\label{eq:rs}
 \mathbf{v}[\mathbf{R}_{m}] =\sum_{n=1}^N \mathbf{T}(\mathbf{R}_{m}-\mathbf{R}_{n};\epsilon) 
  \,\mathbf{g}[\mathbf{R}_{n}]\,.
 \end{equation}
This equation results from inserting 
\begin{equation}
\mathbf{f}(\mathbf{r}') = \sum_{n=1}^{N} \mathbf{g}[\mathbf{R}_{n}] \phi_{\epsilon}(\mathbf{r}' - \mathbf{R}_{n})
\end{equation}
into Eq. \ref{eq:v-field} and defining the ``Regularized Stokeslet" as
\begin{equation}
\label{1}
\mathbf{T}(\mathbf{r};\epsilon) = \int \mathrm{d} \mathbf{r}'\,
\mathbf{T} (\mathbf{r}-\mathbf{r}') \,\phi_\epsilon(\mathbf{r}')\,.
\end{equation}
The ``blob function" $\phi_\epsilon({\mathbf{r}})$ represents a localized envelope over which
forces $\mathbf{g}[\mathbf{R}_{n}]$, at discrete spatial positions, $\mathbf{R}_{n}$, are transmitted to the fluid.  
The blob function must be centered at zero, integrate to 1 and, as $\epsilon \to 0$, 
must  limit to the Dirac delta function $\delta(\mathbf{r})$. We employ 
a Gaussian blob function \cite{camley13}, 
$\phi_\epsilon(\mathbf{r})=\frac{1}{2\pi\epsilon^2} e^{-r^2/(2\epsilon^2)}$ in all of our calculations, but other
choices are certainly possible.  The numerical evaluation of $\mathbf{T}(\mathbf{r};\epsilon)$
for our quasi-2D Saffman-Delbr\"uck systems is described in detail in Ref.
\onlinecite{camley13}.

\begin{figure}
\begin{center}
\includegraphics[scale=0.69]{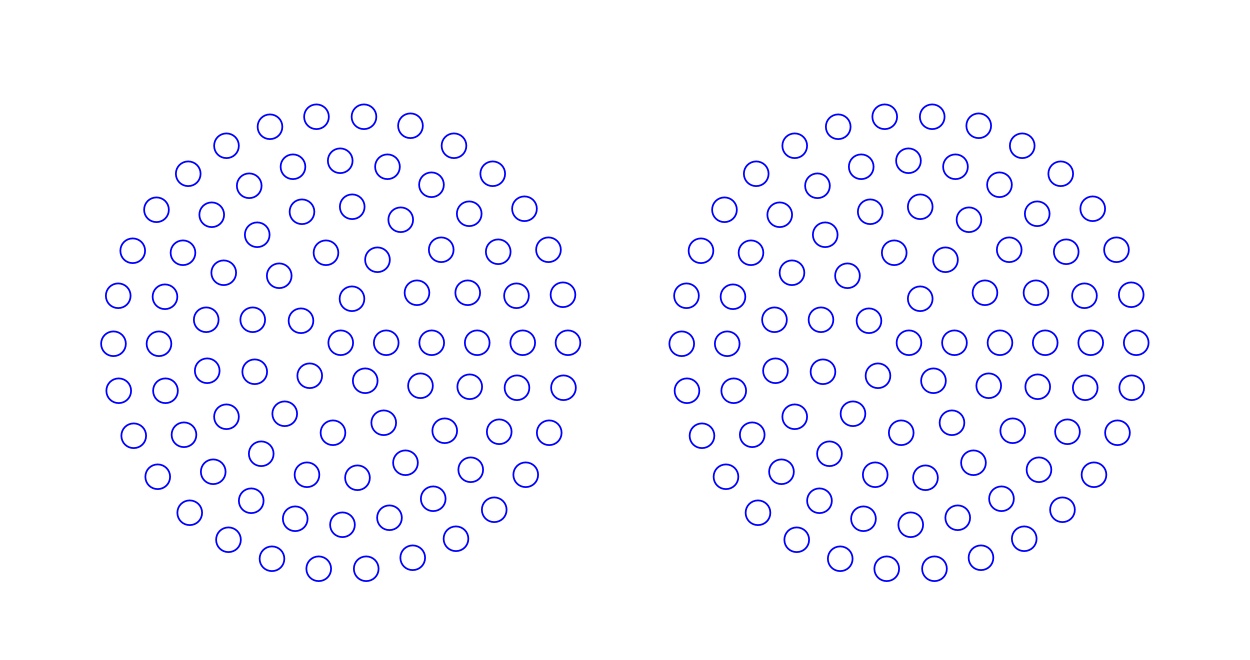}
\caption{\label{fig:circle-blob}
Cartoon of the blob distribution on a pair of disks. In the actual calculations, thousands of blobs
are distributed on the disks. }
\end{center}
\end{figure}

It should be clear that Eq. \ref{eq:rs} is a simple matrix equation.  If we restrict ourselves to only
calculating velocities over the same set of positions where we impose forces, i.e. $\mathbf{R}_{m} \in \{ \mathbf{R}_{n} \}$,
the relevant matrix is square.  We choose the blob positions, $\{\mathbf{R}_{n}\}$, to tile all particles embedded in 
the fluid (see Fig. \ref{fig:circle-blob}) and solve Eq. \ref{eq:rs} to determine the forces, $\{\mathbf{g}[\mathbf{R}_{n}]\}$,
necessary to effect a specified set of velocities, $\{\mathbf{v}[\mathbf{R}_{m}] \}$.  {\em The velocities are always chosen
to describe rigid body motions within each particle} so that $\mathbf{v}[\mathbf{R}_{m_i}]=\mathbf{V}_i+\mathbf{\Omega}_i\times ( \mathbf{R}_{m_i} - \mathbf{R}_{c_i})$ for
each particle $i$.  Here, the subscript $m_i$ is used to index the subset of blobs associated with particle $i$ and $\mathbf{R}_{c_i}$ is the hydrodynamic center of particle $i$.  
We note that blobs are chosen to cover the entire disk,
rather than just the perimeter. This is because, as discussed in Ref.\cite{camley13} 
mandating rigid body motion of the perimeter does not ensure rigid
body motion in the disk interior. In practice, the generalized minimal residual method algorithm (GMRES) \cite{gmres}
is used to solve Eq. \ref{eq:rs} for the blob forces, $\{\mathbf{g}[\mathbf{R}_{n}]\}$.
The blob forces combine to yield the total forces and torques on the individual
particles,
\begin{eqnarray}
\label{eq:force-torque-rs}
\mathbf{F}_i &=& \sum_{m_i} \mathbf{g}[\mathbf{R}_{m_i}]\, \nonumber \\
\mathbf{\tau}_i &=& \sum_{m_i} (\mathbf{R}_{m_i}- \mathbf{R}_{c_i})\times \mathbf{g}[\mathbf{R}_{m_i}]\,.
\end{eqnarray}
Therefore, the RS method provides a route to determining the F-vector corresponding to a given V-vector, using
the language of the preceding section.  The numerical calculations involve a large number of blobs ($N$) to discretize the particles,
but the F-vector and V-vector reflect forces/torques and velocities/angular velocities on the particles and are modestly sized (six elements each for wo
particle calculations).  As a practical matter, it is most convenient to calculate the resistance
matrix by choosing the V-vector input to the RS calculation to be composed entirely of zeros, except for a single element.
The F-vector corresponding to this input then corresponds to a complete column of the resistance matrix.  This process
is repeated for each element of the V-vector to generate the full resistance matrix, column by column.  The diffusion matrix
is then obtained by inversion of the resistance matrix.

For concreteness and to emphasize numerical details we provide a step-by-step procedure for
calculating the diffusion matrix corresponding to the case of two identical disks of radius $r_0$ (see Fig. \ref{fig:topside}) 
embedded within the membrane.
\begin{enumerate}

\item Discretize each disk with $N/2$ blobs, see Fig.~\ref{fig:circle-blob}. 
Similar to Ref.~\citenum{camley13}, the inter-blob spacing, $s$, is chosen to be between 
$0.03-0.09 r_0$. $\epsilon$ is set to be half of the spacing.

\item All the elements of the V-vector are zeroed except for
 $v_{1x}\ne0$, {\it i.e.} the disk 1 moves along the x-axis 
 without rotation and the disk 2 is stationary. 
This V-vector is used as input for the RS routine to obtain the forces and torques on the disks (F-vector).

\item The obtained F-vector is divided by $v_{1x}$
to obtain the first column of the resistance matrix.

\item Repeat steps 2 and 3, keeping a single non-zero element of the V-vector each time 
to obtain the remaining columns of  the resistance matrix. For example, in the fifth run, we set all 
the elements of the V-vector to zero except $\Omega_{1}\ne0$, {\it i.e.} disk 1 
only rotates and the disk 2 is stationary. The resulting F-vector determines the fifth 
column of the friction tensor.

\item Calculate 
the diffusion tensor by numerically inverting the friction tensor, see Eq.~\eqref{eq:diff-fric-tensor}.

\item Repeat steps 1-5 for different values of the spacing
 $s$.   We find that the elements of 
the diffusion matrix change linearly in $s$ and this 
allows us to extrapolate to the infinite resolution limit $s=0$.
The reported results are obtained by this extrapolation.

\end{enumerate}

Step six of the above procedure provides a natural criterion for
establishing the credibility of our numerics.
The mean squared error of the linear fit in step 6 is readily determined for
a given physical situation (i.e. specification of 
$d/r_0$ and $\ell_0/r_0$) by considering $n$ different resolutions: $s_1, s_2, \dots, s_n$ . 
If $y_i$ is the numerical value of an element of the diffusion matrix for the resolution $i$ and $y_{fit}(x)$ is the best-fit line 
to the points $y_1,y_2,\dots,y_n$, the root mean square error reads
\begin{equation}
\label{eq:error}
\mathrm{RMSE}=\sqrt{(1/n)\sum_{i=1}^{n} \big[y_i-y_{fit}(s_i)\big]^2}\,.
\end{equation}

The RMSE for all the ``nonzero" elements of the diffusion matrices which are 
reported in this work are two to three orders of magnitude 
smaller than the extrapolated elements themselves.
However, we point out that the data reported in this paper are restricted to separations $h/r_0>0.1$.  At smaller separations the RMSE 
starts to become large and the linear fit over the data taken at various resolutions is clearly poor;
slight adjustments to the resolution within the range indicated above lead to wild vacillations in the predictions.  This is a clear
indication that the RS scheme at computationally practical resolutions is breaking down.  So, although the RS scheme is
not able to directly probe the small separation limit, we have a clear numerical marker that this is the case.  The numbers reported
in this work are meaningful and we have avoided reporting on any regimes that are problematic.  We do note that the problematic
regime is readily captured via lubrication theory results  and the RS numerics and lubrication 
results share a window of common validity (see App.~\ref{app:lubrication}).

\section{Results for two identical disks}
\label{sec:res}
The preceding sections have detailed the mechanics for calculating the resistance and diffusion
matrices for multiple solid bodies embedded in a membrane.  Here, we apply these methods
to the calculation of these matrices for the specific case of two identical disks embedded
within the bilayer.   As mentioned previously, this case has the advantage of maximal symmetry
and allows us to explore some consequences of the hydrodynamic calculations while keeping
the set of physical and geometric parameters to a minimum.  As we will see, even this simplest of systems
is rather complex to describe.

\subsection{Translational Coupled Diffusion}
\label{subsec:comp-w-haim}
The translational coupling elements of the diffusion matrix quantify correlations
in the thermal Brownian motions of the two disks (see Eq. \ref{eq:fluctuations}).
In the limit of point-like particles,  $D_\mathrm{L}^\mathrm{c}$ and $D_\mathrm{T}^\mathrm{c}$, follow immediately 
from the Green's function of Eq. \ref{eq:mem_green}.  Specifically, if the disks
are separated by distance $d$ along the $x$ axis as diagrammed in fig. \ref{fig:topside},
the limiting forms are provided by
\begin{equation}
\label{eq:mu-c-L}
D_\mathrm{L}^\mathrm{c} \approx k_B T T_{xx}(d \hat{\mathbf{x}}) = \frac{k_BT}{4\eta_\mathrm{m}}\frac{\ell_0}{d} \left[
 H_1(d/\ell_0) - Y_1(d/\ell_0) - \frac{2\ell_0}{\pi d}
 \right]\,,
\end{equation}
and
 \begin{multline}
\label{eq:mu-c-T}
 D_\mathrm{T}^\mathrm{c} \approx k_B T T_{yy}(d \hat{\mathbf{x}}) = \frac{k_BT}{4\eta_\mathrm{m} } \left[
 H_0(d/\ell_0)-\frac{H_1(d/\ell_0)}{d/\ell_0} \right.\\\left.- \frac{1}{2} \left( Y_0(d/\ell_0) - Y_2(d/\ell_0) \right)
 +\frac{2\ell_0^2}{\pi d^2}
 \right] \,. 
\end{multline}
Direct use of these two expressions within the diffusion matrix, would be the membrane analog 
to the Kirkwood approximation for these elements.

Recently \cite{Haim09}, Oppenheimer and Diamant have derived a finite size correction to
eqs. \ref{eq:mu-c-L} and \ref{eq:mu-c-T} in the regime where particle separation is well below
the Saffman Delbruck length (i.e. $d \ll \ell_0$).  For disks of radius $r_0 \ll d \ll \ell_0$ they find
\begin{equation}
 \label{eq:mu-size-corr}
 D^\mathrm{c}_\mathrm{L,T} \approx\frac{k_BT}{4\pi\eta_\mathrm{m}} 
 \left[ \ln(2\ell_0/d) -\gamma_\mathrm{e} \pm \frac{1}{2} \mp \frac{r_0^2}{d^2}\right]\,,
\end{equation}
to leading order in $r_0/d$, where the upper (lower) signs correspond to the longitudinal (transverse) diffusion and 
$\gamma_\mathrm{e}\approx0.5772$ is Euler's constant.  (We mention that the
corresponding formula in Ref.\cite{Haim09} contains a sign error.  A brief derivation
of Eq. \ref{eq:mu-size-corr} can be found in App.~\ref{app:corr} to justify the result
provided here.)  In the limit that $r_0 / d \rightarrow 0$, the last term of Eq. 18 vanishes and the remaining
result is the small $d/\ell_0$ limit of eqs. \ref{eq:mu-c-L} and \ref{eq:mu-c-T}.  It is important to emphasize
the limitation of this formula to the $d \ll \ell_0$ regime.  We are unaware of any comparable analytical 
formula(e) that extend the finite size results beyond this regime.
\begin{figure}
\begin{center}
\includegraphics[scale=0.69]{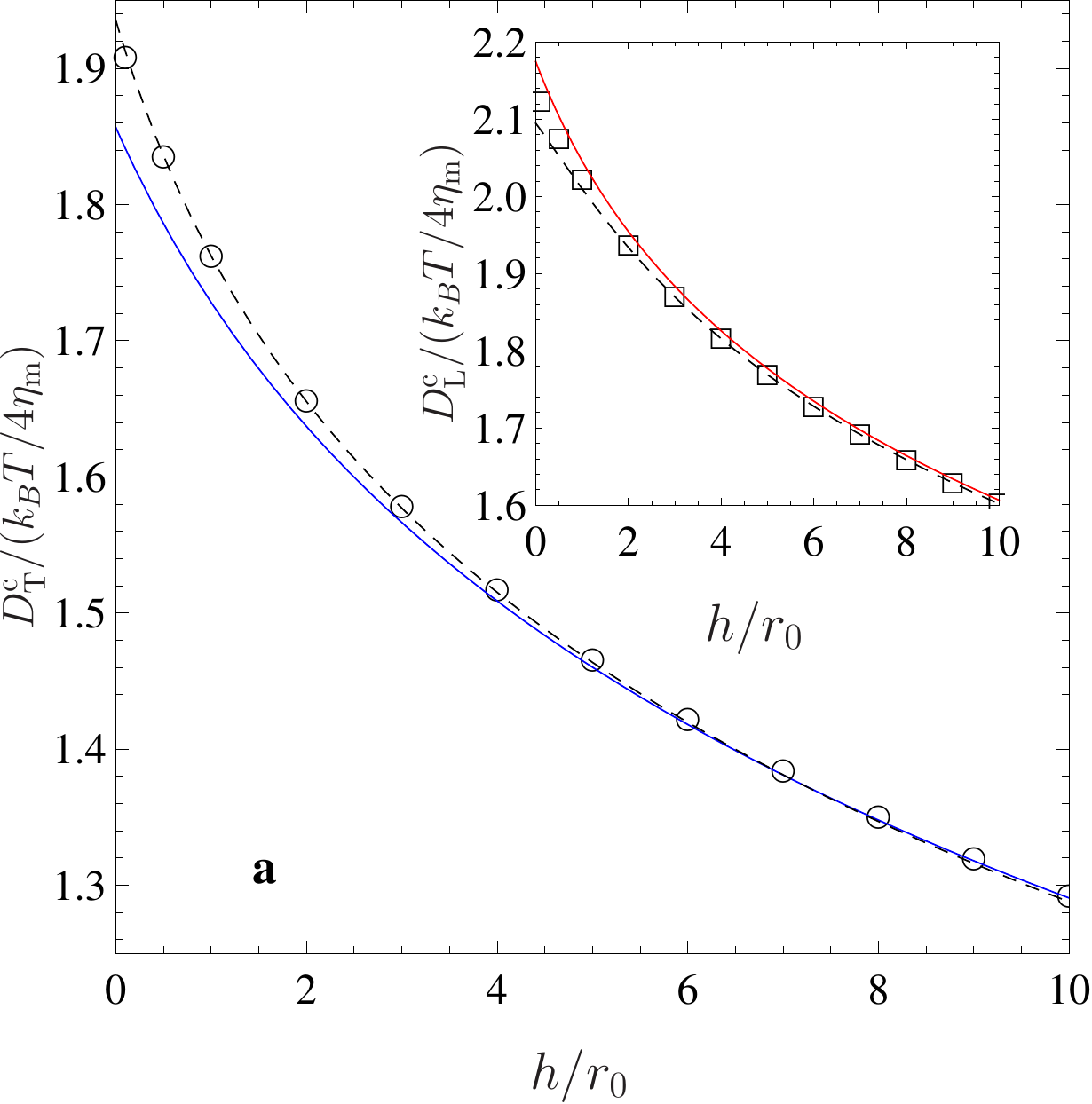}

\vspace{6mm}
\includegraphics[scale=0.7]{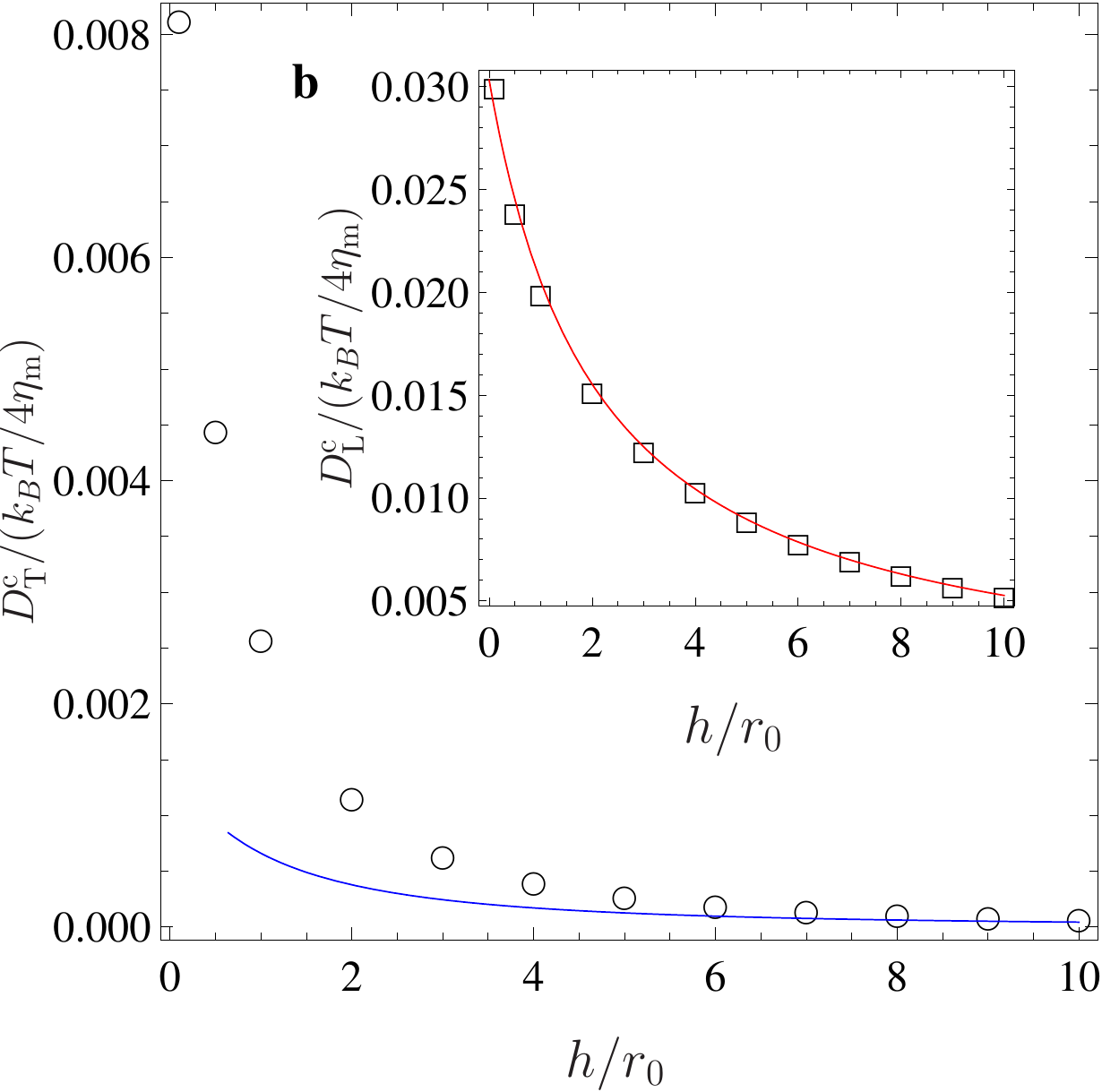}
\caption{\label{fig:HaimShort}
Comparison of the RS data for the coupled-diffusions $D_\mathrm{L}^\mathrm{c}$ and 
$D_\mathrm{T}^\mathrm{c}$ with the related analytical expressions, for 
the Saffman-Delbr\"uck length {\bf (a)} $\ell_0/r_0=10^3$ and 
{\bf (b)} $\ell_0/r_0=0.1$:
The transverse coupled-diffusion $D_\mathrm{T}^\mathrm{c}$ 
scaled by $4\eta_\mathrm{m}/k_BT$ versus the ratio of the 
surface-to-surface distance $h$ and the radius of the disks $r_0$.
The circles are the 
RS data for the transverse coupled-diffusion. The blue and dashed lines 
represent Eqs.~\eqref{eq:mu-c-T} and \eqref{eq:mu-size-corr}, respectively.  As expected, the
dashed line, which includes corrections to account for finite disk size, improves upon the
point particle approximation (blue line).
Inset: the longitudinal coupled-diffusion $D_\mathrm{L}^\mathrm{c}$ 
scaled by $4\eta_\mathrm{m}/k_BT$ 
versus the ratio of the surface-to-surface distance $h$ and the disks' radius $r_0$. 
The squares are the 
RS data for the longitudinal coupled-diffusion. The red lines 
represent Eq.~\eqref{eq:mu-c-L}. 
}
\end{center}
\end{figure}

Results from RS numerical calculations for  $D_\mathrm{L}^\mathrm{c}$ and $D_\mathrm{T}^\mathrm{c}$ 
are displayed in Fig. ~\ref{fig:HaimShort}.  We display two different cases corresponding to the behavior
of small bodies (lipids or proteins, $\ell_0/r_0 =1000$) and large bodies (ten micron scale 
lipid domains, $\ell_0/r_0 = 0.1$) in the  bilayer.  In both cases it is observed that the point-like-particles approximation
performs well until the disks come within several radii  of one another.  For separations of
$h=8 r_0$ and larger, eqs. ~\eqref{eq:mu-c-L} and \eqref{eq:mu-c-T} are indistinguishable from
the full numerical calculations.

There are clear deviations from the point-particle results for small inter-particle separations.  In
the $\ell_0/r_0 \gg 1$ regime these differences are nicely explained by the theory of Oppenheimer
and Diamant (Eq. \ref{eq:mu-size-corr}), though the theory is not quantitatively accurate at very
small separations (as expected).  In the regime of particles large compared to $\ell_0$ we lack
an analytical prediction for finite size corrections to eqs. ~\eqref{eq:mu-c-L} and \eqref{eq:mu-c-T}.
Though finite-particle-size effects make negligible contributions to $D_\mathrm{L}^\mathrm{c}$ for the  $\ell_0/r_0 = 0.1$,
case, there are large effects seen in $D_\mathrm{T}^\mathrm{c}$.  In fact, for 
two identical circular domains with $r_0=10$ microns on a membrane with $\ell_0=1$ micron, at $h=1$ micron separation
the prediction of Eq.~\eqref{eq:mu-c-T} for the transverse mobility is off by more than 800\% from the numerical results.
Although it is perhaps unsurprising that finite particle size should play a major role at such separations, it is interesting
to see just how big the effect is, especially in contrast to the related (lack of) effect on the longitudinal motion.

\subsection{Self-Diffusion}
\label{subsec:change-in-mus}

The self-diffusion matrix elements,
 $D_\mathrm{L}^\mathrm{s}$, $D_\mathrm{T}^\mathrm{s}$,
are affected by the presence of other proximal solid particles
~\cite{montgomery77,montgomeryDeutch77}, though this
influence requires a finite particle size and is a shorter
ranged hydrodynamic effect as compared to that observed
in the coupling diffusion elements.  Indeed, many common theoretical/simulation
tools including the Kirkwood approximation \cite{doi}, Brownian dynamics with 
hydrodynamic interactions \cite{mccammon} and related
methodologies neglect all influence of neighboring particles on self-diffusivity.

In this subsection, we present RS predictions for how
disk self diffusion is impacted by the presence of a second nearby disk. To this end, we compare the self-diffusion of a disk in the two disk system 
with the self-diffusion of an isolated single disk control. 
We consider 
two different scenarios: 
(1) both tracked disk and perturbing disk are freely moving, and (2) 
the tracked disk is free, but the perturber is fixed in place.
We report 
the relative change in the self-diffusions compared to the 
single disk case as
\begin{equation}
 \label{eq:rel-err}
 \Delta D^\mathrm{s}_\mathrm{x} = \frac{D^0_\mathrm{x}-D^\mathrm{s}_\mathrm{x}}{D^0_\mathrm{x}}\,,
\end{equation}
where $D^0_\mathrm{x}$ is the (control) self-diffusion of an isolated single disk, 
and $D^\mathrm{s}_\mathrm{x}$ is the self-diffusion of a disk 
in the two disk system. x= L, T and R, for the 
longitudinal, transverse and rotational mobilities.

\subsubsection{Both disks freely moving}
\label{sec:both_moving_self}

\begin{figure}
\begin{center}
\includegraphics[scale=0.74]{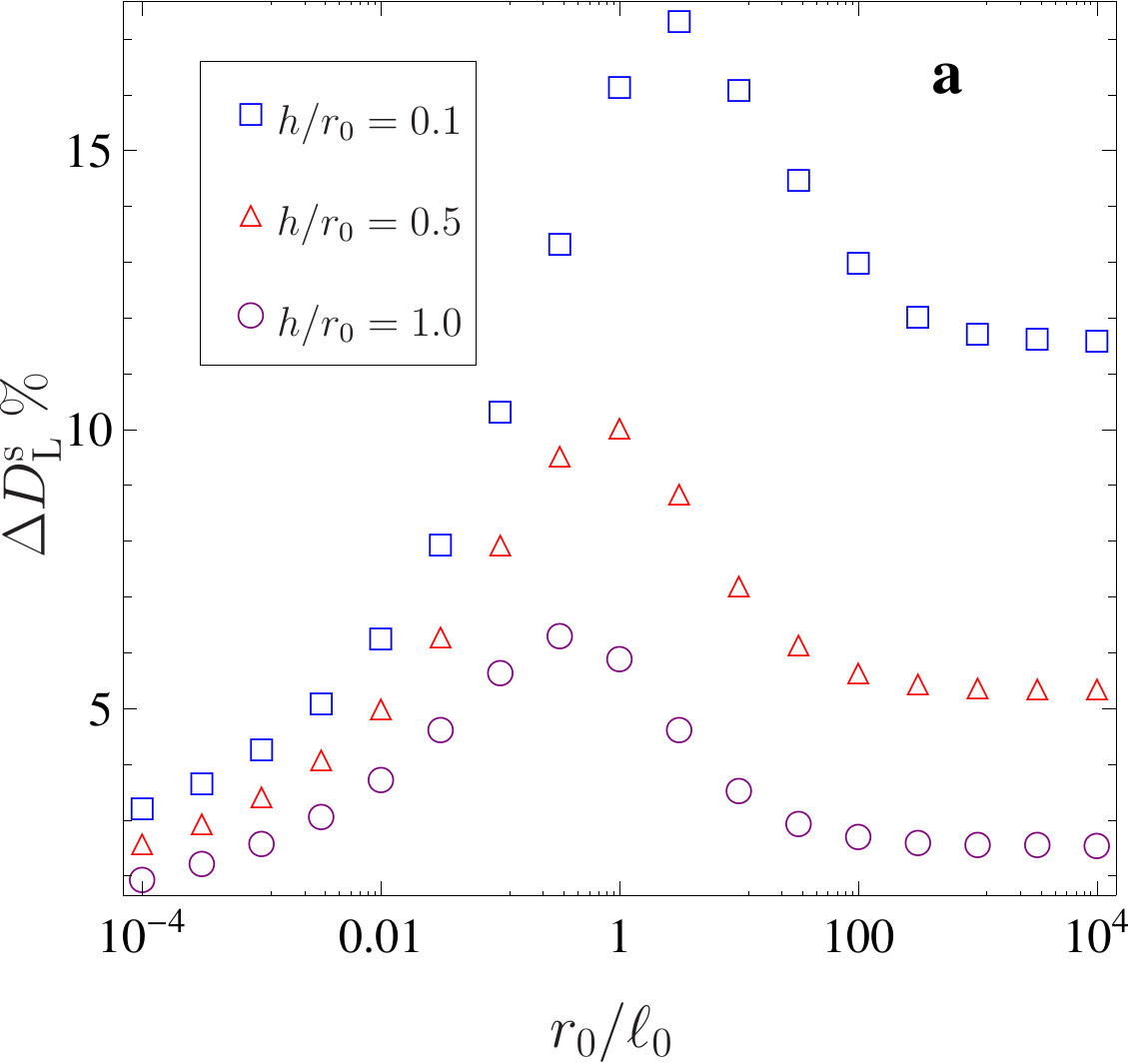}

\vspace{6mm}
\includegraphics[scale=0.73]{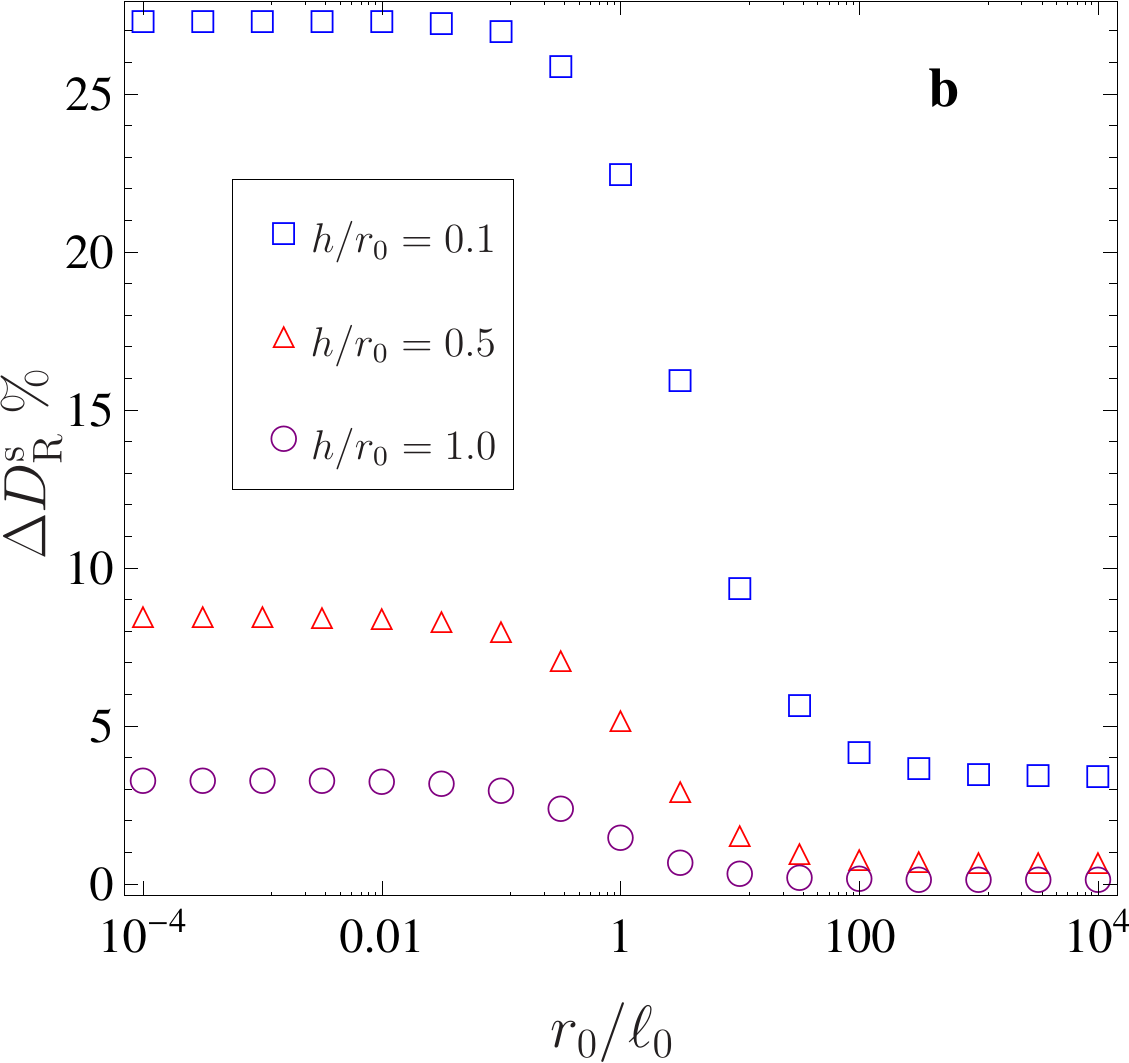}
\caption{\label{fig:self-both-moving} 
Two disks free to move and rotate: the relative change in the 
(a) longitudinal and (b) rotational self-diffusion versus 
the ratio of the radius of the disks $r_0$ and the Saffman-Delbr\"uck length $\ell_0$. 
In both figures (a) and (b) the squares, triangles and circles represent 
the separations $h/r_0=0.1,~0.5$ and 1, respectively.  The percentage changes
to $\Delta D_\mathrm{T}^\mathrm{s}$ are all less than 5\% for the cases considered here
and are not explicitly plotted.
}
\end{center}
\end{figure}

The relative change in longitudinal self-diffusion as a function of 
$r_0/\ell_0$ is depicted in Fig.~\ref{fig:self-both-moving}a.  The
plots display results for three different disk geometries corresponding to 
$h/r_0 = 0.1, 0.5, 1.0$ swept over a wide range of Saffman-Delbr\"uck lengths.
For a given choice of $h/r_0$, any particular choice of $r_0$ (or $h$) yields an identical two-disk geometry
upon proper scaling and it is most useful to regard changes along horizontal axis ($r_0/\ell_0$) in this figure 
as resulting from altering the membrane $\ell_0$ while maintaining a given two-disk geometry.

The first point to note about Fig. \ref{fig:self-both-moving}a is that the cases plotted cover a significantly
abbreviated range of values for $h/r_0$ relative to the cases considered in the context of coupled diffusion.  The passive
effect of the second disk on the motion of the first is a considerably shorter ranged phenomenon than the
direct transmission of force from disk one to disk two at play in the coupling diffusion coefficients.    At edge-to-edge
separations equal to one disk radius, the strongest observed influence on self diffusion is only on the order of 5\% and
shrinks still further for larger separations.  Only for very closely placed disks does the effect become appreciable.  We
also note that the extreme values of $\ell_0$ occurring at the edges of the plot place the system in either a limiting 3D hydrodynamic regime independent of $\ell_0$
(at small $\ell_0$) or the regularized 2D limit of the Saffman-Delbr\"uck model (at large $\ell_0$) with weak (logarithmic \cite{saf}, see Eq. \ref{eq:mu-size-corr})
dependence on $\ell_0$.  
The relative change in the transverse self-diffusion $\Delta D_\mathrm{T}^\mathrm{s}$ 
remains very small at all separations, although it 
displays the same trends as in Fig.~\ref{fig:self-both-moving}a. 
The maximum relative change in the transverse self-diffusion 
for the separation ratio $h/r_0=0.1$ is less than 5\% and we have
not explicitly included $\Delta D_\mathrm{T}^\mathrm{s}$ plots.

\begin{figure}
\begin{center}
\includegraphics[scale=0.75]{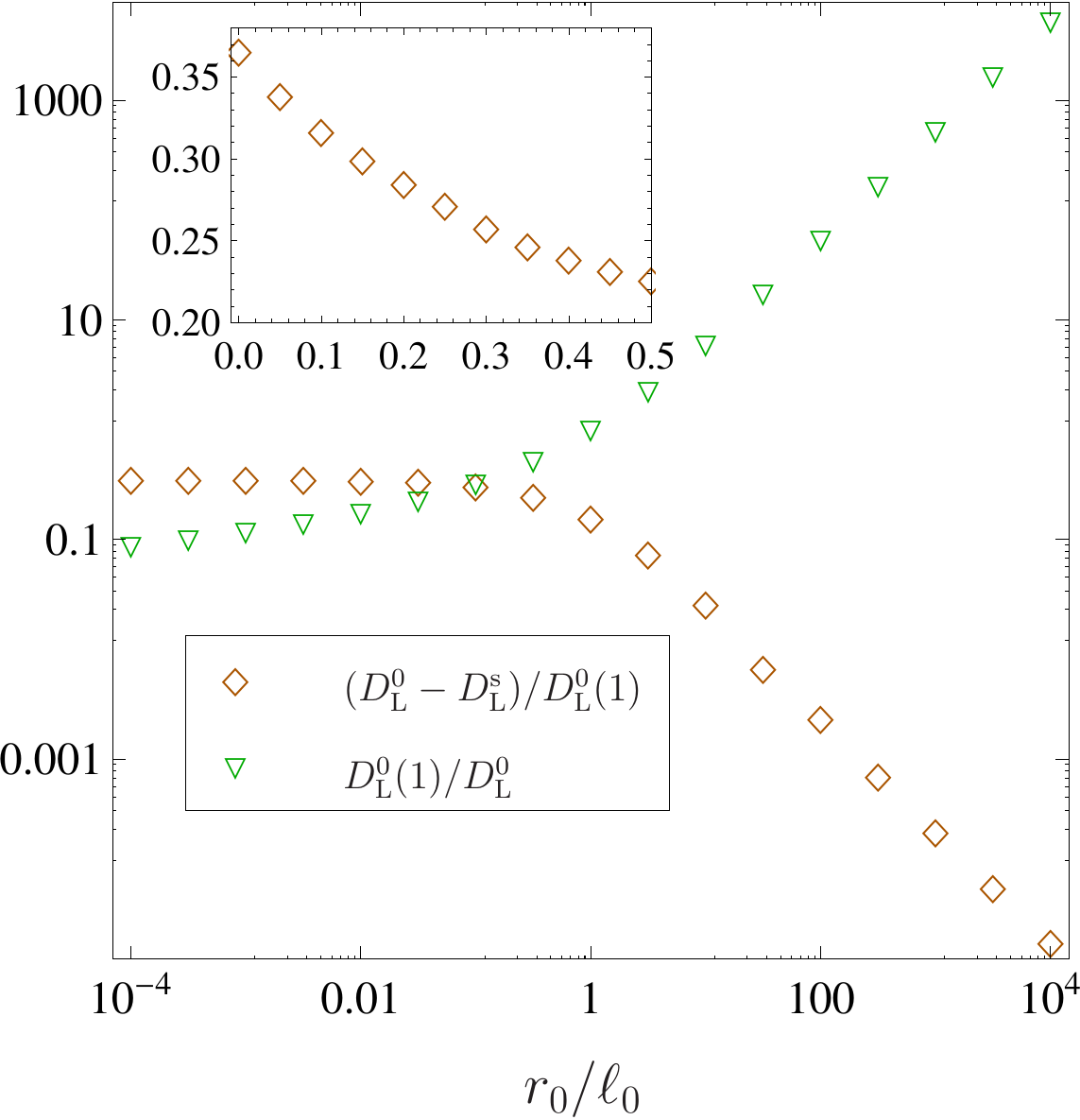}
\caption{\label{fig:why-pick} 
Two disks free to move and rotate: Diamonds show the dimensionless difference between the 
longitudinal self-diffusions of a disk in the presence of another one $(h/r_0=0.1)$ and a single disk 
$(D^0_\mathrm{L}-D^\mathrm{s}_\mathrm{L})/D^0_\mathrm{L}(1)$ versus the ratio of the 
radius of the disks and the Saffman-Delbr\"uck length $r_0/\ell_0$. 
Down triangles represent the dimensionless inverse of the longitudinal self-diffusion of the single disk $D^0_\mathrm{L}(1)/D^0_\mathrm{L}$
versus $r_0/\ell_0$. $D^0_\mathrm{L}(1)$ is the longitudinal self-diffusion coefficient of 
the single disk for $\ell_0/r_0=1$. 
Inset: Short and intermediate distance close-up of $(D^0_\mathrm{L}-D^\mathrm{s}_\mathrm{L})/D^0_\mathrm{L}(1)$ on the linear scale. 
}
\end{center}
\end{figure}

The most striking feature of Fig. \ref{fig:self-both-moving} is the
non-monotonic behavior for the longitudinal translation case, with
a maximum for disks of radius comparable to the 
Saffman-Delbr\"uck length, {\it i.e.} $r_0 \sim \ell_0$.  Although
this feature draws the eye, we do not believe it to be as
interesting as might first be believed.  We recall that the
quantities plotted in Fig. \ref{fig:self-both-moving} are 
{\em relative} differences in self-diffusion,  with the difference itself
plotted in ratio to the corresponding self-diffusion of an isolated
single-disk control.  Both the numerator and denominator of the ratio
display strong, but clearly monotonic dependence on $r_0 / \ell_0$ in the vicinity
of $r_0 / \ell_0$ (see Fig. \ref{fig:why-pick}) .  The observed maximum simply results 
from the product of two functions, one sharply increasing and the other sharply decreasing.
The appearance of a maximum can be qualitatively reproduced using less elaborate
numerical models, as is explicitly demonstrated in App.~\ref{app:dim-mon} using a Kirkwood approximation
for a simplified system.


The relative changes in rotational self-diffusion are presented in Fig.~\ref{fig:self-both-moving}b.
The trends of $\Delta D_\mathrm{R}^\mathrm{s}$ are completely different than the $\Delta D_\mathrm{L}^\mathrm{s}$ case,
as might have been anticipated due to the shorter ranged hydrodynamic flows associated with rotational motion
relative to translational motion.  Here, the larger effects are seen in the 2D hydrodynamic regime relative to the
3D regime and the crossover is completely monotonic.  Unlike translational diffusion, rotational diffusion does not
suffer from a Stokes paradox in 2D and requires no regularization by a surrounding subphase to obtain finite results \cite{saf}; the large
$\ell_0$ limit for rotational diffusion saturates to a true 2D limit without any residual logarithmic corrections.  

In reconciling the 
somewhat backward trends between panes (a) and (b) of Fig. \ref{fig:self-both-moving} it is helpful to recognize that the effects
described in this figure result from the perturbation to single disk motion caused by constraining a nearby region of the membrane to undergo
only rigid-body motion.  We expect that the largest changes to self diffusion will occur when the unperturbed flows around a moving isolated disk 1 deviate 
most strongly from rigid body flows at the location of disk 2.  In the limit of large $\ell_0$ in Fig. \ref{fig:self-both-moving}a the velocity field proximal to a single translating disk is very nearly constant.
The constraint of rigid body motion over the envelope of disk 2 is hardly a constraint at all and has minimal effects on the translational self diffusion.  Rotational motions,
on the other hand, generate considerably shorter ranged flows which vary appreciably proximal to the rotating disk and a strong effect is seen
in the large $\ell_0$ limit of Fig. \ref{fig:self-both-moving}b.  In the small $\ell_0$ (3D) regime the translational flows are not as flat as in the opposite regime,
which explains the relatively larger changes on the right of Fig. \ref{fig:self-both-moving} a.  The rotational flows are also more rapidly varying in the 3D regime
relative to the 2D regime, and actually die off so rapidly away from the rotating disk that they are quite small in the vicinity of the second disk, explaining the small effects on the right.
of Fig. \ref{fig:self-both-moving} b.

\subsubsection{One Disk Stationary, the Other Free to Move and Rotate}
\label{sec:fixed_disk}

In biological contexts, it is common for proteins and larger assemblies to be immobilized
on the membrane surface due to interactions with cytoskeletal filaments and related cellular
structures \cite{lodish}. It is thus natural to consider the influence of immobilized perturbers
on particle self diffusion in addition to the case considered in the previous section. 
The self-diffusion of a disk adjacent to a completely immobile (i.e. no translation and
no rotation) neighboring disk is a distinct hydrodynamic problem from that considered
in the previous section.  Indeed, constraining $\mathbf{v}=0$ in a region proximal 
to the diffuser is clearly a significantly more severe restriction than only requiring rigid-body 
motion over the same region.  Within the RS framework, immobility of disk 2 is
enforced by setting $v_{2x}=v_{2y}=\Omega_2=0$ in Eq. \ref{eq:2-obj-fric-tensor-general}.
Further, we only require the $f_{1x}$, $f_{1y}$ and $\tau_1$ components of the F-vector
to determine the mobilities of disk 1 as the forces on the immobilized disk are irrelevant to us.  
Eq.\ref{eq:2-obj-fric-tensor-general} is thus reduced to an effective single particle resistance
problem
\begin{eqnarray}
  \label{eq:1-obj-fric-tensor-general}
\left(
\begin{array}{c}
 f_{1x} \\ f_{1y} \\  \tau_1 
 \end{array}
\right)
&=&
\left(  
  \begin{array}{ccc}
\zeta^\mathrm{11}_\mathrm{xx} & \zeta^\mathrm{11}_\mathrm{xy} &  \zeta^\mathrm{11}_\mathrm{xR} \\
\zeta^\mathrm{11}_\mathrm{yx} & \zeta^\mathrm{11}_\mathrm{yy} &  \zeta^\mathrm{11}_\mathrm{yR} \\
\zeta^\mathrm{11}_\mathrm{xR} & \zeta^\mathrm{11}_\mathrm{yR} & \zeta^\mathrm{11}_\mathrm{R}
\end{array}
  \right)
\left(
\begin{array}{c}
 v_{1x} \\ v_{1y} \\  \Omega_1 \end{array}
\right) \nonumber \\
& = & \left(  
  \begin{array}{ccc}
\zeta^s_L & 0 & 0 \\
0 & \zeta^s_T &  \zeta^s_{RT} \\
0 & \zeta^s_{RT} & \zeta^s_{R}
\end{array}
  \right)
\left(
\begin{array}{c}
 v_{1x} \\ v_{1y} \\  \Omega_1 \end{array}
\right) 
\, .
\end{eqnarray}
The first of these equations corresponds to the case of two general objects with object 2 immobilized;
the second equation is specific to the symmetries present in the two disk case with the second disk immobilized.  We stress that the individual elements of
the above matrices are identical to those appearing in Eqs. \ref{eq:2-obj-fric-tensor-general} and \ref{eq:circ-r-mat}
and need to be calculated via the full 2 body RS scheme described in section \ref{sec:rs-method}.  No new 
RS calculations are needed here, beyond those made to calculate the elements of the general two-body resistance
matrix.  The effect of second body immobilization on the self-diffusion of body one is captured through the different
matrix inversion required in this case relative to the case considered in the previous section and as indicated in Eq. \ref{eq:diff-fric-tensor}.
The self-diffusivities of particle 1 in the present case are obtained by inverting the truncated $3\times 3$ resistance matrix
of Eq. \ref{eq:1-obj-fric-tensor-general} and not the full $6 \times 6$ resistance matrix.

\begin{figure}
\begin{center}
\includegraphics[scale=0.63]{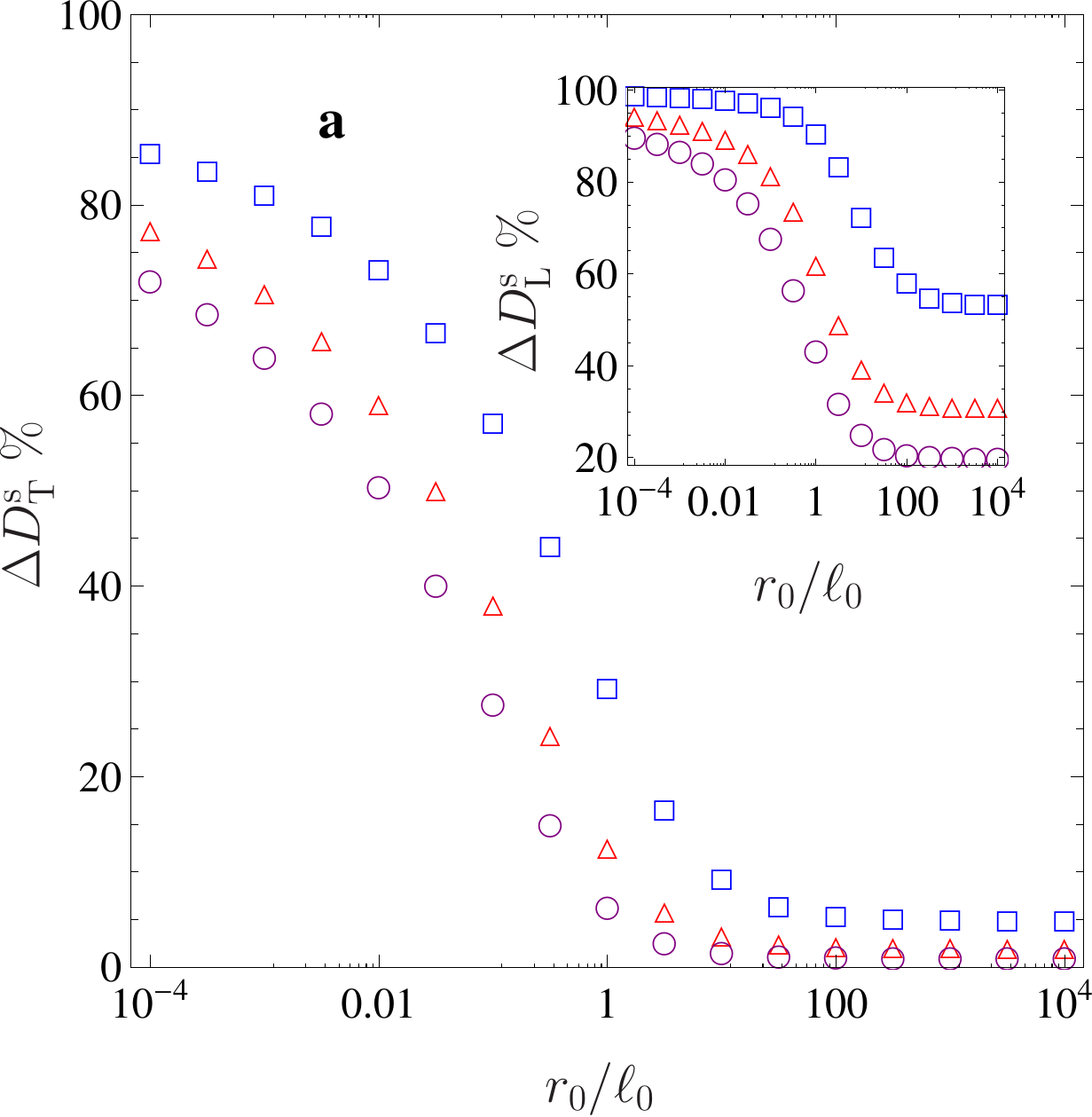}
\vspace{5mm}

\includegraphics[scale=0.64]{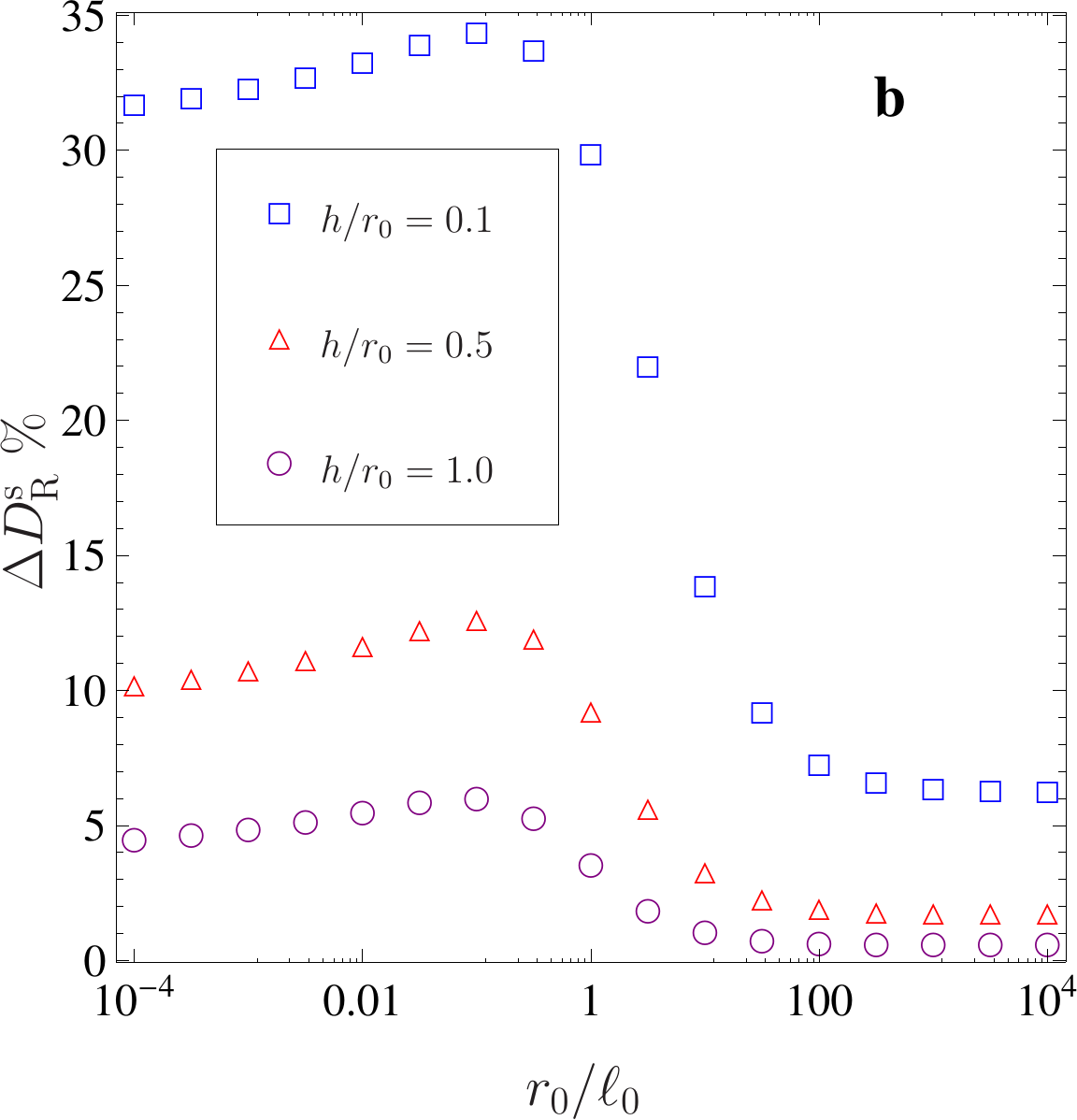}
\caption{\label{fig:self-one-stat} 
An immobile and a free disk (the free disk is free to move and rotate): 
The relative change in the (a) transverse and (b) rotational self-mobilities 
versus the ratio of the radius of the disks $r_0$ and the Saffman-Delbr\"uck length $\ell_0$. 
In both figures (a) and (b) the squares, triangles and circles represent 
the separations $h/r_0=0.1,~0.5$ and 1, respectively. 
Inset: The relative change in the longitudinal self-diffusion.
}
\end{center}
\end{figure}

As expected, compared to a mobile adjacent disk, a stationary disk exerts a considerably stronger 
influence on translational self-diffusions 
,$\Delta D^\mathrm{s}_\mathrm{L,T}$ ( see Fig.~\ref{fig:self-one-stat}a).  
The effect is particularly strong in the 2D regime of large $\ell_0$.  Although
enforcing rigid body motion over the perturbing disk has little effect due to the
nearly homogeneous velocity field around the diffuser at large $\ell_0$, forcing complete immobility
of the perturbing disk has a huge effect and can slow diffusion by nearly a factor
of two.  This is true for both longitudinal and transverse motions.  In the 3D regime, 
flows die off more rapidly away from the diffuser and the perturbation by disk 2 is smaller than 
in the 2D regime.  However, the effect of an immobile perturber is always stronger than a mobile
one; forcing immobility on the fluid flow in a region in space is a stronger perturbation than just
requiring rigid body motion over the same region.

The relative change in the rotational self-diffusion, $\Delta D_\mathrm{R}^\mathrm{s}$, 
is presented in Fig.~\ref{fig:self-one-stat}b. 
Comparing Figs.~\ref{fig:self-both-moving}b and ~\ref{fig:self-one-stat}b, 
 the rotational self-diffusion is affected by the immobile disk 
between 5 \% to 10\% more strongly than the free disks case, but the trends
are similar.

\section{Linked Dimers: Rotating vs Non-Rotating Monomers}
\label{sec:dimers}

\begin{figure}
\begin{center}
\includegraphics[scale=0.45]{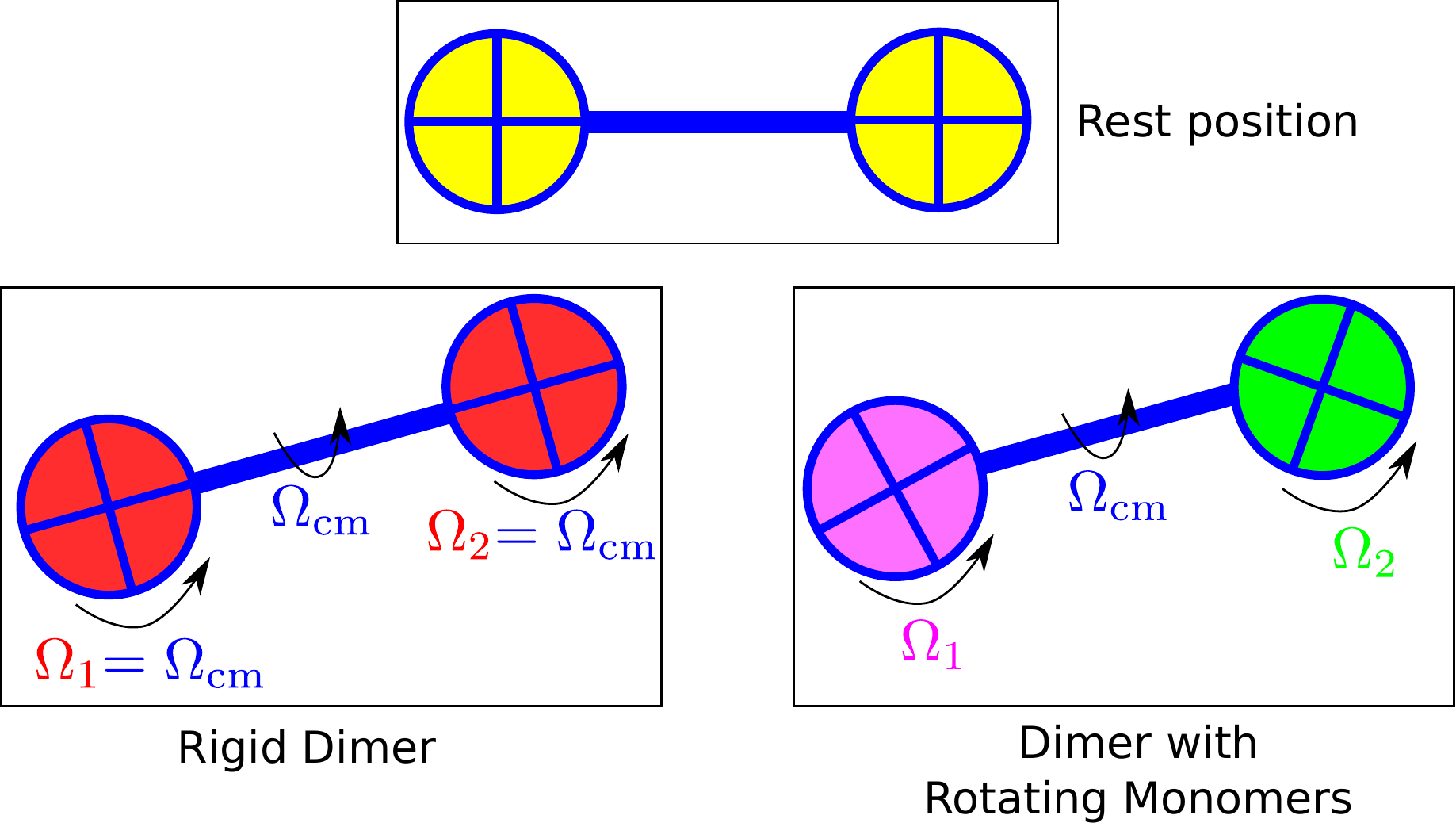}
\caption{\label{fig:dimerss} 
Two disk dimers are derived from the general two disk systems of the previous sections under the constraint of constant separation
between the disks.  ``Rigid dimers" are further constrained to treat the dimer assembly as a single rigid body. ``Dimers with rotating monomers''
allow the monomers to freely spin with angular velocities that are not slaved to the overall rotation of the dimer about its center of mass.  }
\end{center}
\end{figure}

Inspired by the recent experiments of Knight and coworkers
\cite{knight10}, we consider the motion of two-disk dimers (Fig. \ref{fig:dimerss}) on the
membrane surface.
The experimental systems involve constructs of two lipid associated proteins bound
together by molecular (peptide) linkers of various lengths. (Trimers were also studied 
in the experiments, but are not considered here.)  In a previous study \cite{camley13},
we treated these dimers as a completely rigid body and modeled the translational diffusion
for direct comparison to experiment.  We did not consider rotational diffusion in that work because rotation 
was not tracked experimentally and we had not developed the theoretical methodology to treat any 
motions beyond rigid body translations and rotations.  Given the flexible nature of the peptide linkers
used in experiment, the assumption of a completely rigid dimer is questionable and might naively
be expected to be particularly problematic for rotational motions.  Requiring the individual 
monomers of the dimer to rotate with the same angular velocity as the whole assembly is a significant
constraint.  The purpose of this section is to evaluate how relaxing this constraint affects overall dimer rotation.

As we have seen in Sec. \ref{sec:fixed_disk}, the calculation of particle mobilities/diffusivities
under the constraint that certain motions are disallowed is straightforward.
Once the elements of the resistance matrix have been calculated for the general case absent constraints,
the hard work has been done.  Imposing the constraint(s) simply means choosing input V-vectors consistent
with the constraint(s) and only tracking those component(s) of the F-vector orthogonal to constrained motions.
The reduced resistance matrix obtained in this manner may then be inverted to obtain the diffusion matrix for
the unconstrained motions.  This procedure was particularly simple for the example of a completely immobile 
monomer studied in Sec. \ref{sec:fixed_disk}, since our starting basis involved single disk motions and the 
reduced basis was obtained simply by zeroing out half of the V-vector corresponding to motions of disk 2.
Similarly, the forces/torque associated with disk 2 were ignored, which was also completely natural within
the single disk basis.

The present case involving dimer motions is slightly more complicated because our resistance matrix elements
have been calculated in the single disk basis, but we need to impose constraints that involve both disks.  These constraints
are handled much more naturally in a basis of center-of-mass and relative motions of the two disks.  To this end,
we consider a V-vector
\begin{equation}
\label{eq:rigid-dimer-Uv}
\left(
\begin{array}{c} 
v_{\mathrm{cm},x} \\
v_{\mathrm{cm},y} \\
\Omega_{\mathrm{cm}} \\
\Delta\Omega_1 \\
\Delta\Omega_2 \\
v_{\mathrm{rel}} 
\end{array}
\right)
=
\mathbf{T}\,\cdot
\left(
\begin{array}{c} 
v_{1x} \\
v_{1y} \\
v_{2x} \\
v_{2y} \\
\Omega_1 \\
\Omega_2 
\end{array}
\right)
\end{equation}
defined by the transformation matrix
\begin{equation}
\label{eq:Uv-rigid}
\mathbf{T}=\left(
  \begin{array}{cccccc}
\frac{1}{2} & 0 & \frac{1}{2} & 0 & 0 & 0 \\
0 & \frac{1}{2} & 0 & \frac{1}{2} & 0 & 0 \\
0 & -\frac{1}{d}  & 0 & \frac{1}{d}  & 0 & 0 \\
0 & \frac{1}{d} & 0 & -\frac{1}{d}  & 1 & 0 \\
0 & \frac{1}{d} & 0 & -\frac{1}{d} & 0 & 1 \\
 -1 & 0  & 1 & 0 & 0 & 0\\
\end{array}
  \right).
\end{equation}
$v_{\mathrm{cm},x}$, $v_{\mathrm{cm},y}$ and $\Omega_{\mathrm{cm}}$ are the translational and angular velocities of the
dimer relative to its center of mass. $v_{\mathrm{rel}}$ is the relative velocity of the monomers along the axis joining them and
$\Delta \Omega_{1,2}$ are the angular velocities of the individual disks measured relative to $\Omega_{\mathrm{cm}}$.  The
dimers we consider will always have a fixed separation and $v_{\mathrm{rel}}=0$.  Additionally, ``rigid dimers" are defined
by $\Delta \Omega_{1,2}=0$, whereas dimers with freely rotating monomers impose no such constraint; see Fig. \ref{fig:dimerss}.
If we desire to preserve a similar expression of the fluctuation-dissipation relationship in our new basis as seen in the original
two-particle basis (i.e. Eq. \ref{eq:fluctuations}), it is essential to define the elements of our transformed F-vector as
\begin{equation}
\label{eq:rigid-dimer-Uf}
\left(
\begin{array}{c} 
f_{\mathrm{cm},x} \\
f_{\mathrm{cm},y} \\
\tau_\mathrm{cm}\\
\Delta\tau_1 \\
\Delta\tau_2 \\
f_{rel}  \\
\end{array}
\right)
=(\mathbf{T}^\intercal)^{-1}\, \cdot
\left(
\begin{array}{c} 
f_{1x} \\
f_{1y} \\
f_{2x} \\
f_{2y} \\
\tau_1 \\
\tau_2 
\end{array}
\right)
\end{equation}
with the inverse of the transpose of $\mathbf{T}$ given explicitly by 
\begin{equation}
  \label{eq:Uf-rigid}
(\mathbf{T}^{-1})^{\intercal}=(\mathbf{T}^\intercal)^{-1}=\left(
  \begin{array}{cccccc}
1 & 0 & 1 & 0 & 0 & 0 \\
0 & 1 & 0 & 1 & 0 & 0 \\
0 & -\frac{d}{2} & 0 & \frac{d}{2} & 1 & 1 \\
0 & 0& 0 & 0 & 1 & 0 \\
0 &0 & 0 & 0 & 0 & 1 \\
-\frac{1}{2} & 0  &\frac{1}{2} & 0 & 0 & 0 \\
\end{array}
  \right)\,.
\end{equation}
It then follows that the resistance and mobility matrices in our new basis are related
to their counterparts in the two-particle basis as
\begin{eqnarray}
\boldsymbol{\zeta}' & = &(\mathbf{T}^\intercal)^{-1} \boldsymbol{\zeta} \mathbf{T}^{-1} \nonumber \\
\mathbf{D}' &=& \mathbf{T} \mathbf{D} \mathbf{T}^\intercal. \label{eq:transformed_matrices}
\end{eqnarray}
Further, the matrices obey the expected inverse relation to one another, $\mathbf{D}' = k_B T \boldsymbol{\zeta}'^{-1}$
and the elements of $\mathbf{D}'$ are simply related to the transformed coordinate displacements in a short time interval in analogy to Eq.
\ref{eq:fluctuations}: $2D'_{\alpha \beta} \Delta t = \langle\Delta q'_\alpha \Delta q'_\beta \rangle$.

We calculate the diffusion coefficients for the rigid dimer by considering only the diagonal upper left $3 \times 3$ corner of the $\boldsymbol{\zeta}'$
matrix defined by Eq. \ref{eq:transformed_matrices} and inverting this reduced resistance matrix to obtain a diagonal $3 \times 3$ diffusion matrix.  The
resulting diffusion coefficients $D_L$ (longitudinal translational motion along the dimer axis), $D_T$ (transverse translational motion perpendicular to the
dimer axis) and $D_{rot}$ (rotational motion of the dimer) have the explicit forms
 \begin{eqnarray}
\label{eq:rigid-dimer-diffusion}
D_\mathrm{L}/(k_BT) &=& \frac{1}{2(\zeta^\mathrm{s}_\mathrm{xx}+\zeta^\mathrm{c}_\mathrm{xx})} \,\,\,\, \mbox{(rigid dimer results)} \nonumber\\
D_\mathrm{T}/(k_BT) &=& \frac{1}{2(\zeta^\mathrm{s}_\mathrm{yy}+\zeta^\mathrm{c}_\mathrm{yy})}  \\ 
D_\mathrm{R}/(k_BT)&=&\frac{1}
{2(\zeta^\mathrm{s}_\mathrm{r}+\zeta^\mathrm{c}_\mathrm{r}) 
    -2 d (\zeta^\mathrm{s}_\mathrm{rt}+\zeta^\mathrm{c}_\mathrm{rt})
    + \frac{d^2}{2} (\zeta^\mathrm{s}_\mathrm{yy}-\zeta^\mathrm{c}_\mathrm{yy})} \nonumber \,.
\end{eqnarray}
It was verified that results obtained in this manner are numerically identical to direct RS calculations treating
the entire dimer as a single rigid body (as in Ref. \cite{camley13}).  The corresponding quantities for the dimer with rotating monomers 
are calculated by considering the upper left $5 \times 5$ corner of $\boldsymbol{\zeta}'$
and performing the $5 \times 5$ inversion to obtain the diffusion matrix.  This matrix is not diagonal, however the first three elements 
along the diagonal still correspond to mean square displacements and rotation of the dimer about the center of mass.  The results are
\begin{eqnarray}
\label{eq:mobil-non-rigid-mobil}
D_\mathrm{L}/(k_BT) &=& \frac{1}{2(\zeta_\mathrm{xx}^\mathrm{c}+\zeta_\mathrm{xx}^\mathrm{s})} \,\,\,\, \mbox{(rotating monomers results)} \nonumber \\
  D_\mathrm{T}/(k_BT)&=& \frac{\zeta_\mathrm{r}^\mathrm{c}-\zeta_\mathrm{r}^\mathrm{s}}
  {2\big((\zeta_\mathrm{rt}^\mathrm{c}-\zeta_\mathrm{rt}^\mathrm{s})^2+
  (\zeta_\mathrm{r}^\mathrm{c}-\zeta_\mathrm{r}^\mathrm{s})
  (\zeta_\mathrm{yy}^\mathrm{c}+\zeta_\mathrm{yy}^\mathrm{s})
  \big)} \,, \\
D_\mathrm{R}/(k_BT)&=& \frac{2(\zeta_\mathrm{r}^\mathrm{c}+\zeta_\mathrm{r}^\mathrm{s})}
  {d^2\big(
  (\zeta_\mathrm{r}^\mathrm{c}+\zeta_\mathrm{r}^\mathrm{s})
  (\zeta_\mathrm{yy}^\mathrm{s}-\zeta_\mathrm{yy}^\mathrm{c})
  -(\zeta_\mathrm{rt}^\mathrm{c}+\zeta_\mathrm{rt}^\mathrm{s})^2 
  \big)} \,.\nonumber
\end{eqnarray}

Our interest is in the comparison between rotational diffusion for the two different dimer assemblies.
To this end, we compare the rotational diffusion of a rigid dimer with fixed monomers $\widetilde{D}_\mathrm{R}^\mathrm{fixed}$ and 
a dimer with rotating monomers $\widetilde{D}_\mathrm{R}^\mathrm{rot}$ via the relative difference in the rotational self-diffusions as
\begin{equation}
\label{eq:rel-dif-drot}
\Delta \widetilde{D}_\mathrm{R}=\frac{\widetilde{D}^\mathrm{rot}_\mathrm{R}-\widetilde{D}^\mathrm{fixed}_\mathrm{R}}{\widetilde{D}^\mathrm{rot}_\mathrm{R}}\,.
\end{equation}
\begin{figure}
\begin{center}
\includegraphics[scale=0.9]{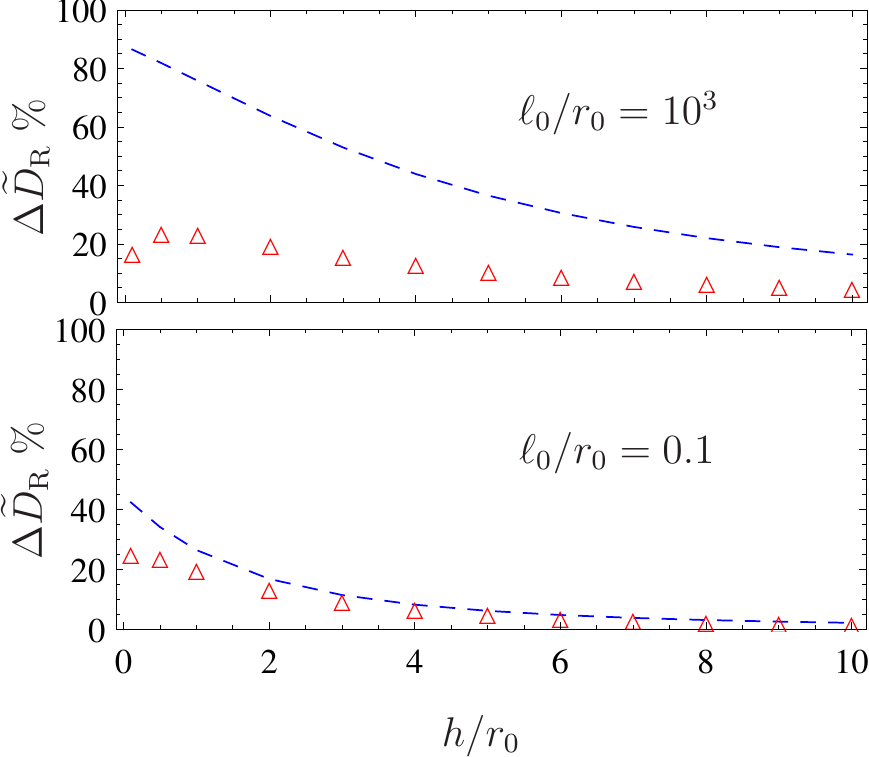}
\caption{\label{fig:dimer} 
Relative difference between rotational self-diffusions for dimers with and without freely rotating monomers 
(calculated by Eq.~\eqref{eq:rel-dif-drot}) versus the ratio of the 
distance of the closest approach between the monomers $h$ and the radius of the monomers $r_0$. The upper and lower plots are 
for the Saffman-Delbr\"uck lengths $\ell_0/r_0 = 10^3$ and 0.1, respectively.
The triangles correspond to the full calculation including all
hydrodynamic interactions between the monomers.  The dashed curves represent the
results obtained by neglecting monomer-monomer hydrodynamic interactions (see text).}
\end{center}
\end{figure}
Figure~\ref{fig:dimer} plots $\Delta \widetilde{D}_\mathrm{R}$ 
as a function of the dimer separation for two Saffman-Delbr\"uck lengths: 
$\ell_0/r_0= 10^3$ (protein-sized monomers) and 0.1 (domain-sized monomers). 
The results for the two regimes are rather similar, with only modest differences between
the two models.  However, it is interesting to note that this similarity between the two regimes
as well as the small magnitudes seen in the plots both result from particle-particle hydrodynamic
interactions damping otherwise more significant effects.  The dashed lines in fig. \ref{fig:dimer}
result from approximating the full calculation by turning off all hydrodynamic interactions between
the monomers (i.e. by zeroing all off-diagonal contributions to the resistance matrix and calculating
all diagonal elements in the absence of the second disk in Eq. \ref{eq:circ-r-mat} prior to transforming
to $\boldsymbol{\zeta}'$).  In this limit, Eq. \ref{eq:rel-dif-drot} takes the simple form
$\Delta \widetilde{D}_\mathrm{R} = 1/(1 + \frac{d^2 \zeta_{xx}^{s0}}{4 \zeta_{r}^{s0}})$, where the ``$s0$"
superscript indicates the self resistances are calculated for an isolated single particle. 

It is expected that $\Delta \widetilde{D}_\mathrm{R}$ is a positive quantity.  Constraining the monomers
to rotate with the overall rotation of the dimer can only lead to additional dissipation beyond what the
unconstrained system would experience under a prescribed angular velocity.  As the dashed lines in fig. \ref{fig:dimer}
indicate, this effect is potentially large (especially for protein sized monomers) when calculated in
the absence of direct monomer-monomer hydrodynamic coupling.  It is expensive to force the monomers
to rotate.  However, this cost is substantially reduced when the full calculation is performed.  With hydrodynamic
coupling present, the monomers naturally spin in the same direction as the dimer rotations, so forcing the
completely in-phase motion between dimer and monomers is less costly.  The effect is especially prominent
for dimers in the 2D hydrodynamic regime, where the hydrodynamic interactions are strong.

\section{An example of noncircular bodies: diamond shaped domains}
\label{sec:diamonds}
\begin{figure}
\begin{center}
\includegraphics[scale=0.65]{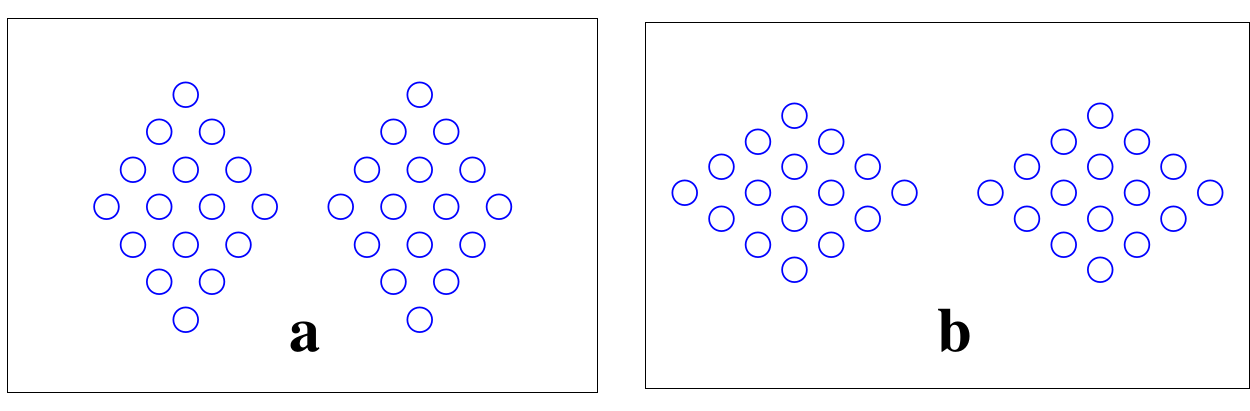}
\caption{\label{fig:diamonds-blob} 
Cartoon of the blob distribution for the 
(a) side-to-side and (b) tip-to-tip diamonds. In the calculations,
thousands of blobs tile each diamond.
}
\end{center}
\end{figure}

Though the preceding examples have been limited to circular disks
to maximize symmetry and simplify analysis, a strength of the RS methodology is the ability to consider bodies of arbitrary shape.
As a specific example, we consider the diamond-shaped 
lipid domains studied in Ref. \cite{petrov12}.  Previously, we used the RS technique
to study single diamond-shaped domains \cite{camley13}.  This section considers the diffusion
of diamond domains in proximity to other diamond shaped domains, which
corresponds to the actual experimental conditions of Ref. \cite{petrov12}.
In these experiments, micron-scale diamond-shaped (with aspect ratio of 1.42, see Fig. \ref{fig:diamonds-blob}) solid lipid domains in a
background fluid lipid phase (with Saffman-Delbr\"uck length $\ell_0 \sim 1 \mbox{$\mu$m}$) were tracked in time to measure the translational and rotational diffusion coefficients
of the domains.  Though precise area coverages of the diamonds are not reported for the experiments, the provided images clearly display diamonds in proximity to one another
and it is natural to ask how the diffusive characteristics of a tracked domain are expected to be altered by a neighbors.  The results of Sec. \ref{sec:both_moving_self}
for disks suggest that that since the diamond size is close to $\ell_0$ in these experiments, the effect should be nearly as strong as is possible.

To determine the impact of hydrodynamic interactions on the 
self-diffusion of the diamonds, 
RS calculations were performed for diamond pairs on a 
membrane with the Saffman-Delbr\"uck length of 1 micron (see 
the discussion in Ref.~\cite{camley13} regarding the estimation of $\ell_0$ in the diamonds experiment).
In accordance with typical experimental numbers \cite{petrov12}, the short axis of the 
diamonds is chosen to be  2.00\,$\mu m$ and the long axis 2.84\,$\mu m$,
corresponding to the aspect ratio of 1.42 mentioned above.  This coincides with
an effective radius $r_0\approx 0.95\mu m$, for direct comparison to the data
in Ref. \cite{petrov12}.   (The effective radius is the radius of a hypothetical disk
that shares the same area as the diamond.)

\begin{figure}
\begin{center}
\includegraphics[scale=0.64]{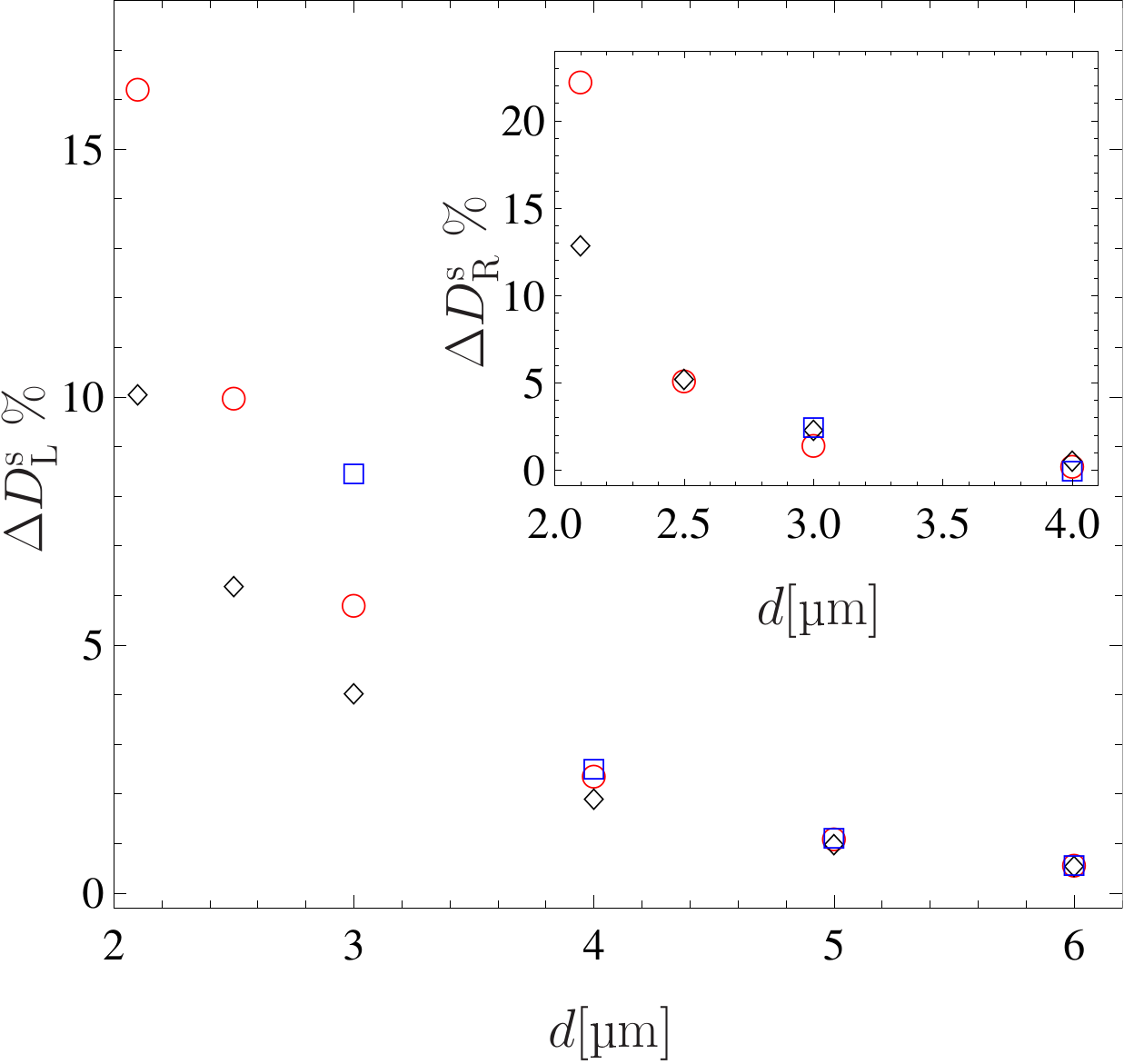}
\caption{\label{fig:diamond} 
Two identical diamonds free to move and rotate: the relative change in the 
longitudinal and rotational (inset) self-diffusions versus the separation between the 
diamonds, for $\ell_0 = 1\mu m$\,. The aspect ratio of the diamonds 
is 1.42 and the shorter diagonal is 2\,$\mu m$ 
(equivalent to the effective radius of $r_0\approx 0.95\mu m$\,).  $d$ is the 
separation between diamond centers.
The black diamonds and blue squares respectively represent the data for the 
side-to-side and tip-to-tip geometries, 
see Fig.~\ref{fig:diamonds-blob}. The circles show the data for the disks with the radius $r_0=0.95~\mu m$ 
 on the same membrane.  Note that the shortest separations are excluded for the tip-to-tip geometry as they
correspond to overlapping configurations.
}
\end{center}
\end{figure}

Figure~\ref{fig:diamond} plots the relative change (as defined in Eq. \ref{eq:rel-err}) in the longitudinal and rotational self-diffusions
for side-to-side and tip-to-tip diamond geometries in addition to the corresponding quantities obtained by approximating the diamonds
as effective circles with areas identical to the diamond areas. 
The relative change depends both on the separation and orientation of the  diamonds.  For 
separations in the range $2\mu m \lesssim d \lesssim3\mu m$ 
the relative change is between 5\% to 25\% and dies off rapidly with increasing separation
(especially for rotational motion).  Although closely spaced neighboring domains can quantitatively
affect the measured diffusion, the effect is not large and is not strikingly different from what could
be estimated by approximating the domains as circles.  In particular, domain-domain
hydrodynamic interactions probably can not explain the apparent experimental inconsistencies 
between rotational and translational domain motion discussed at some length in Ref. \cite{camley13}.

\section{Conclusion}
\label{sec:summary}

The RS approach \cite{cortez01,cortez05} is a versatile and powerful tool for studying
particulate flows in traditional 3D hydrodynamic geometries and, more recently, in 
the quasi-2D geometries associated with lipid bilayer membranes \cite{camley13}.
This work has further validated the membrane RS methodology of Ref. \cite{camley13} and
demonstrated its utility in application to problems involving multiple membrane-embedded
objects.  The applications considered herein involve comparisons to analytical theory and
experimental measurements, but one of the most promising future applications of this
work will be its use in parameterizing and validating approximate simulation schemes for the 
dynamics of protein laden membranes and related inhomogeneous biomembrane systems.

Standard techniques for simulating particulate flows in 3D \cite{mccammon,stokesian} start with
a pairwise approximation to the diffusion matrix for the full many body system. Typically, the
Rotne-Prager approximation \cite{rotne_prager} is employed.  In the case of membrane systems,
there is no Rotne-Prager analog available.  One possibility for developing membrane simulations
analogous to ``Brownian Dynamics with Hydrodynamic Interactions" \cite{mccammon} is to
numerically tabulate the coupling elements of the diffusion matrix as a function of particle
separation to build up a pairwise approximation to the diffusion matrix that can be implemented
in simulations, as has been suggested for 3D systems \cite{wang2013assessing}.  The RS calculations presented in this work are a natural method to use for such
a tabulation and could, perhaps, be supplemented with the lubrication results of Ref. \cite{bussel92}
to capture near-field effects, as in Stokesian dynamics \cite{stokesian} simulations in 3D.  

An alternate strategy for studying protein laden membranes based on an immersed boundary (IB)
scheme \cite{peskin} has recently been introduced \cite{camley14}.  This method does not
require input of a diffusion matrix, but rather tracks fluid flow explicitly and derives hydrodynamic
interactions from the fluid velocity field.  However, the treatment of particles in this scheme
is approximate (as in any IB simulation) and has only been verified to agree
with the far-field predictions related to those of Oppenheimer and Diamant \cite{Haim09}.  Comparison
between RS results and the membrane IB simulations will be an important future test of
this simulation methodology and may suggest future improvements to this (or other) 
biomembrane simulation scheme incorporating hydrodynamic interactions.  Work along these
lines is currently being pursued.

\begin{acknowledgments}
F.B. thanks Haim Diamant and Gilad Haran for helpful conversations. 
This research was supported in part by a grant from the BSF (Grant No. 2012084 to FB) and 
the NSF (Grant No. CHE-1153096 to FB).
\end{acknowledgments}

\appendix

\section{Comparison of the RS numerics with the lubrication approximation}
\label{app:lubrication}

\begin{figure*}[ht]
\begin{center}
\includegraphics[width=7.0in]{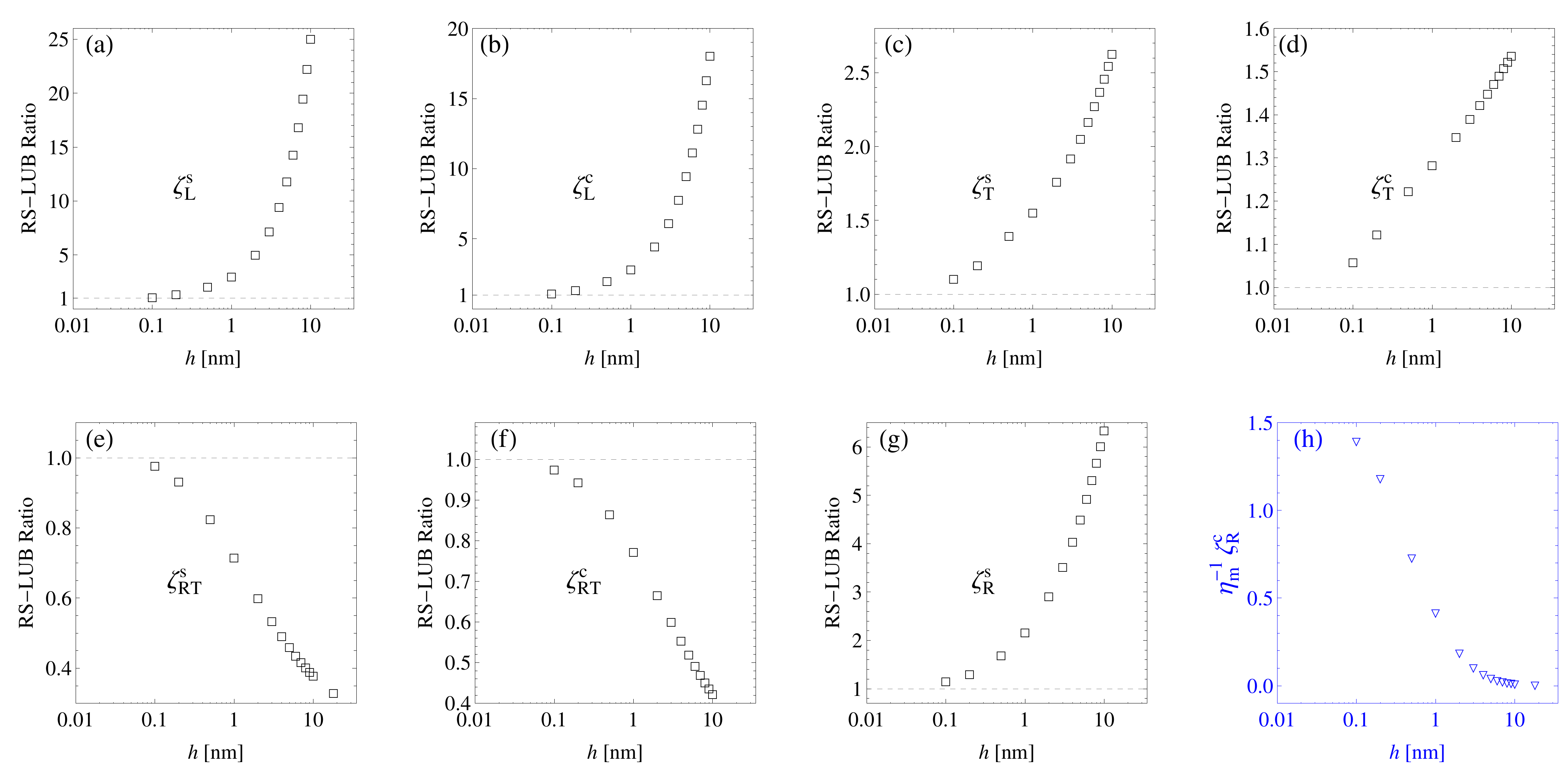}
\caption{\label{fig:lubr} 
RS numerics and the lubrication (LUB) limit for two identical circular membrane proteins: 
(a)-(g): The ratio of RS to LUB elements of the friction tensor versus the surface-to-surface separation $h$. 
The radius of the proteins is $r_0=1$ nm and the 
Saffman-Delbr\"uck length is $\ell_0=1~\upmu$m. The printed friction tensor element on each figure 
shows that the comparison is made for that specific element. (h): The correlated rotation friction tensor element as a function of 
separation.
}
\end{center}
\end{figure*}

This appendix compares RS numerical results with the limiting small-separation lubrication (LUB) analytical expressions of Bussell, Koch and Hammer~\cite{bussel92}. 
The LUB results for each element of the friction tensor 
are presented in Table~\ref{tab:lub}.  
RS calculations as described in the main text were used to compute the friction tensor for two identical proteins with radii 
$r_0=1$~nm and the surface-to-surface separation $h$ 
ranging from 0.1~nm to 10~nm embedded in a membrane with the Saffman-Delbr\"uck length 1 micron.
The ratio of each element of the RS friction tensor to the corresponding LUB friction tensor element is plotted in Fig.~\ref{fig:lubr}.

\begin{center}
\begin{table}
\caption{\label{tab:lub} Lubrication predictions for two disks of radius $r_0$ with closest separation distance $h$ \cite{bussel92}.  These are the limiting results in the regime of small $h/r_0$, where both lengths are also much smaller than $\ell_0$.  The result ``$O(1)$" indicates saturation to a constant at
small separations in contrast to the divergent results expected for the remaining elements.}
\begin{tabular}{|cc|cc|}
\hline
Friction Tensor element &&& Lubrication prediction
\\ \hline
$\eta_\mathrm{m}^{-1}\zeta_\mathrm{L}^\mathrm{s}$  &&& $\frac{3}{2}\pi\,(h/r_0)^{-3/2}$
\\ \hline
$\eta_\mathrm{m}^{-1}\zeta_\mathrm{L}^\mathrm{c}$  &&& $-\frac{3}{2}\pi\,(h/r_0)^{-3/2}$
\\ \hline
$\eta_\mathrm{m}^{-1}\zeta_\mathrm{T}^\mathrm{s}$  &&& $\pi\,(h/r_0)^{-1/2}$
\\ \hline
$\eta_\mathrm{m}^{-1}\zeta_\mathrm{T}^\mathrm{c}$  &&& $-\pi\,(h/r_0)^{-1/2}$
\\ \hline
$\eta_\mathrm{m}^{-1}r_0^{-1}\zeta_\mathrm{RT}^\mathrm{s}$  &&& $\pi\,(h/r_0)^{-1/2}$
\\ \hline
$\eta_\mathrm{m}^{-1} r_0^{-1}\zeta_\mathrm{RT}^\mathrm{c}$  &&& $\pi\,(h/r_0)^{-1/2}$
\\ \hline
$\eta_\mathrm{m}^{-1} r_0^{-2}\zeta_\mathrm{R}^\mathrm{s}$  &&& $2\pi\,(h/r_0)^{-1/2}$
\\ \hline
$\eta_\mathrm{m}^{-1} r_0^{-2}\zeta_\mathrm{R}^\mathrm{c}$  &&& $O(1)$\\
\hline
\end{tabular}
\end{table}
\end{center}

Panes ~\ref{fig:lubr}~(a)~and~(b) show that RS numerics and LUB predictions for the longitudinal resistance tensor elements fully converge to one another by
$h=0.1$~nm. 
Figures~\ref{fig:lubr}~(c)-(g) show near convergence between RS and LUB results by $h=0.1$~nm for other elements of the friction tensor . 
As discussed in the main text, smaller $h$ geometries begin to show numerical artifacts in the RS computations for computationally 
feasible blob resolutions, which is why calculations were halted at $h=0.1$~nm.  The coupled rotational 
friction element (pane ~\ref{fig:lubr}~(h) ) lacks a limiting LUB result to compare against, and we have plotted the friction element itself in this case.  The
results are in good agreement with the numerics of Ref.~\cite{bussel92}. 

\section{Finite Size Correction to the Coupled-Diffusions}
\label{app:corr}

In this appendix, we outline the derivation of finite-size corrections to the 
coupled-diffusions $D_\mathrm{L}^\mathrm{c}$ and $D_\mathrm{T}^\mathrm{c}$, as 
originally presented in Ref.~\cite{Haim09}.  The derivation proceeds as follows:

\begin{enumerate}
\item Two disks of radii 
$r_1$ and $r_2$, separated by the distance $d$ considered.  It
is assumed that $r_1,r_2 < d \ll \ell_0$ and, without loss of generality, that
the position of disk 1 is the origin and disk 2 lies on the x axis.

\item Since both disk radii and separation are all far smaller than $\ell_0$, the
flow profile generated around a forced disk 1 in the vicinity of disk 2, but ignoring the presence of disk 2, 
can be approximated by the formula for the flow around a forced disk in 2D:
\begin{eqnarray}
\label{eq:near-field}
v_i(\mathbf{r}) &\approx& \frac{1}{4 \pi \eta_{m}} \bigg[ -\left\{\ln (r/2\ell_0) + \gamma_e + 1/2 \right\}\delta_{ij} + \frac{r_i r_j}{r^2}  \bigg . \nonumber \\
 &+ & \bigg . \frac{r_1^2}{2 r^2} 
 \left(\delta_{ij}-\frac{2r_i r_j}{r^2} \right)\bigg]f_{1,j}\,.
 \end{eqnarray}
 Here, $\gamma_e$ is the Euler-Mascheroni constant, $f_{1,j}$ are the 
 cartesian components of force on disk 1 and the Einstein summation
 convention is assumed.

\item The membrane Fax\'en law derived by Oppenheimer and Diamant \cite{Haim09}
relates the components of force applied to disk 2, $f_{2,j}$, the components of velocity of 
disk 2, $U_{2,j}$ and the ambient membrane velocity field were particle 2 not present  {\em assuming
the 2D hydrodynamic regime specified in point 1}.  This
field is provided by $v_i$ from Eq. \ref{eq:near-field} in the case of a flow driven 
by forcing on particle 1.  The result is
\begin{equation}
\label{eq:faxen}
\mathbf{f}_2\approx \frac{4\pi\eta_\mathrm{m}}{-\ln(2\ell_0/r_2)+\gamma_\mathrm{e}} 
 \left(
 \mathbf{v} +\frac{r_2^2}{4} \nabla^2 \mathbf{v}
 -\mathbf{U}_2
 \right)\,.
\end{equation}
Under the assumption that particle 2 is unforced, this implies
\begin{equation}
\label{eq:U2}
U_{2i} = v_i + \left(\frac{r_2}{2}\right)^2 \nabla^2 v_i\,
\end{equation}
where $v_i$ and its Laplacian are evaluated at the center
of disk 2, i.e. at $\mathbf{r}=(d,0)$.  Carrying out the 
indicated derivatives and retaining terms only up to order second
order in $r_1/d$ and $r_2/d$ yields
\begin{eqnarray}
U_{2,x} & = & \frac{1}{4 \pi \eta_{m}} \bigg[ -\left\{\ln (r/2\ell_0) + \gamma_e - 1/2 \right\}   - \frac{r_1^2 + r_2^2}{2d^2} \bigg] f_{1,x} \nonumber \\
U_{2,y} & = &  \frac{1}{4 \pi \eta_{m}} \bigg[ -\left\{\ln (r/2\ell_0) + \gamma_e +1/2 \right\}   + \frac{r_1^2 + r_2^2}{2d^2} \bigg] f_{1,y}. \nonumber
\end{eqnarray}

\item Given the geometry specified in point 1 and comparing to
Eq. \ref{eq:circ-r-mat} we have  $D_\mathrm{L}^\mathrm{c}/(k_BT)=U_{2,x}/f_{1,x}$ and 
$D_\mathrm{T}^\mathrm{c}/(k_BT)=U_{2,y}/f_{1,y}$.  This yields the result quoted in
Eq. \ref{eq:mu-size-corr}.
\end{enumerate}

\section{Relative change in the longitudinal self-diffusion:  a dimer and monomer in 
the Kirkwood approximation}
\label{app:dim-mon}

\begin{figure}
\begin{center}
\includegraphics[scale=0.65]{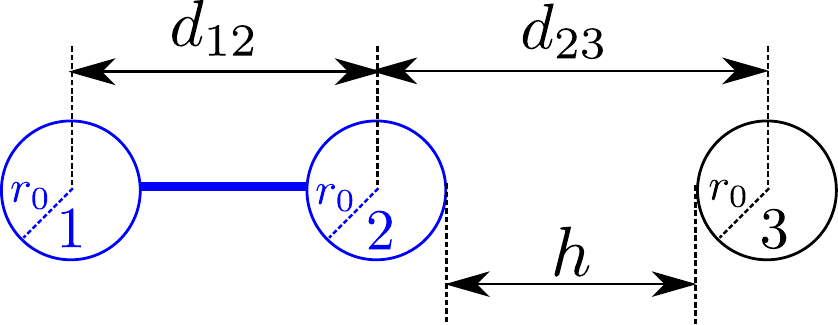}
\caption{\label{fig:dim-mon} A co-linear dimer and monomer.
}
\end{center}
\end{figure}

The non-monotonic behavior observed in Fig. \ref{fig:self-both-moving} 
can be qualitatively captured with a simple 3-bead Kirkwood type model.
The perturbation of a nearby solid disk on the self-diffusivity of a tracked disk 
is due to the rigid-body constraint on fluid flow over the spatial envelope of the
perturber.  For longitudinal motions, we can mimic this full rigid body constraint by
 a simpler rigid-body constraint between two beads.  We consider
the situation pictured in Fig.~\ref{fig:dim-mon}: a diffusing bead is proximal
to a rigid dimer of beads and all three beads are co-linear.
All beads are identical with radius $r_0$ and 
the dimer separation is $d_{12}$.  The distance between the dimer beads and the
monomer are $d_{23}$ and $d_{13}$ with $d_{13}>d_{23}$. $h$, the point of closest
approach between dimer and monomer is given by $h = d_{23} - 2r_0$. 
Within the Kirkwood approximation \cite{doi} the diffusion tensor of the three-bead system for
motions along the x-axis reads
\begin{equation}
 \label{}
\mathbf{D}=
 \left(
 \begin{array}{ccc}
  D^\mathrm{0} & D_\mathrm{L}^\mathrm{c}(d_{12}/\ell_0) & D_\mathrm{L}^\mathrm{c}(d_{13}/\ell_0) \\
  D_\mathrm{L}^\mathrm{c}(d_{12}/\ell_0) & D^\mathrm{0} & D_\mathrm{L}^\mathrm{c}(d_{23}/\ell_0) \\
  D_\mathrm{L}^\mathrm{c}(d_{13}/\ell_0) & D_\mathrm{L}^\mathrm{c}(d_{23}/\ell_0) & D^\mathrm{0} 
 \end{array}
\right)\,,
\end{equation}
where $D^\mathrm{0}$ is the diffusion coefficient for an isolated single disk of radius $r_0$ and $D^\mathrm{c}_\mathrm{L}$ is the longitudinal
 coupled-diffusion in the point-particle approximation given by Eq.~\eqref{eq:mu-c-L}.  $D^\mathrm{0}$ is given by the Hughes, Pailthorpe, White solution \cite{white81}, which we
 implement numerically via the empirical fit introduced by Petrov and Schwille \cite{petrov08}.

\begin{figure}
\begin{center}
\includegraphics[scale=0.65]{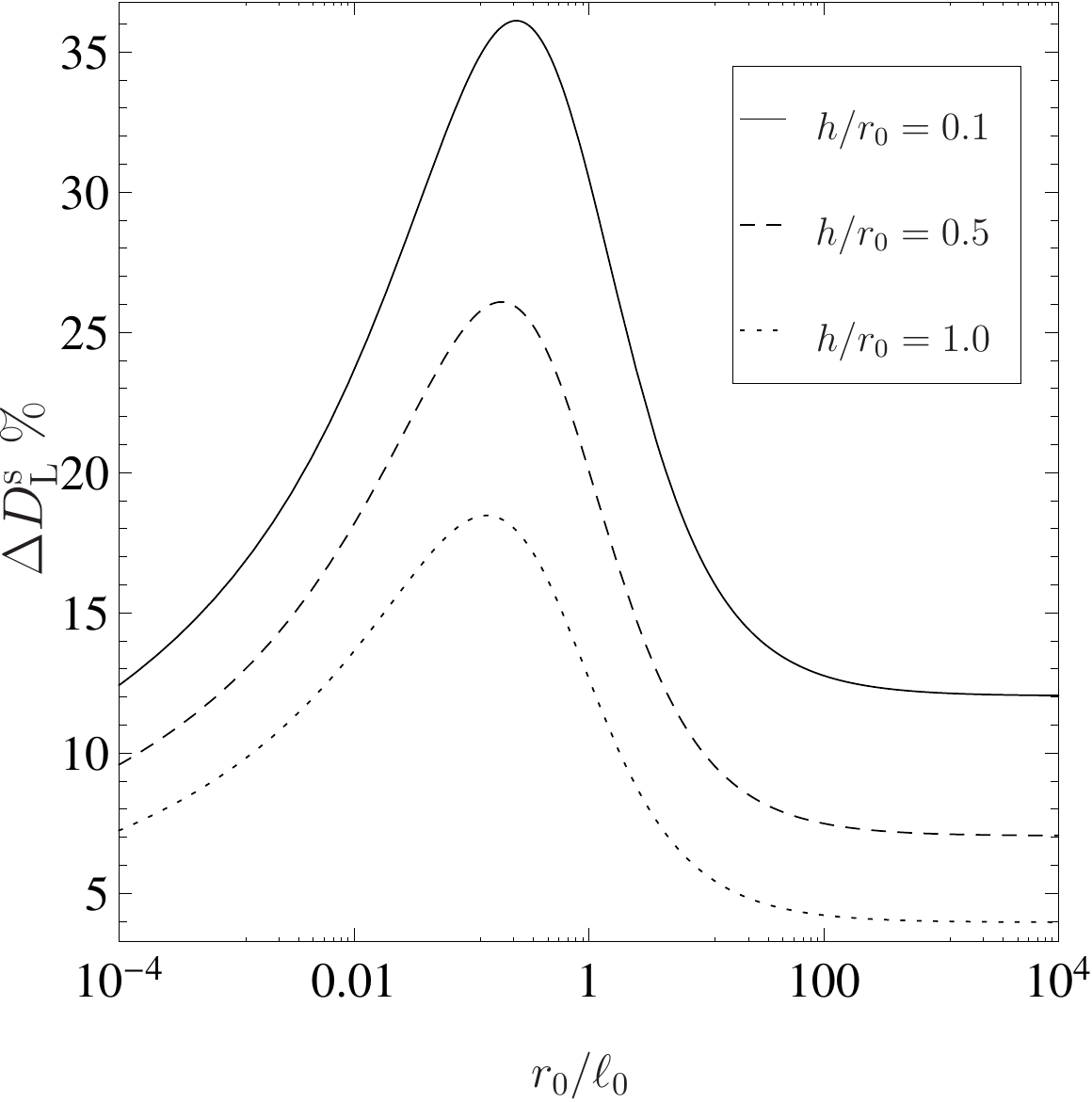}
\caption{\label{fig:dimon}
A dimer and a monomer: the relative change in the self-diffusions as a function of $\ell_0$ 
for three separations.  The plot assumes $d_{12} = 2r_0$ and the geometry
specified in Fig. \ref{fig:dim-mon}.
}
\end{center}
\end{figure}

We assume that a force $f_{3x}$ is applied to the monomer along the x-axis and the dimer is force-free. 
An undetermined Lagrange multiplier $\lambda$ is introduced to enforce the constant separation of beads 1 and 2 in the dimer.  The
resulting mobility problem is thus
\begin{equation}
 \label{eq:dif-eq}
 \left(
 \begin{array}{c}
  v_{1x} \\ v_{2x} \\v_{3x}
 \end{array}
\right)
=
\frac{\mathbf{D}}{k_BT}
\cdot
\left(
 \begin{array}{c}
  -\lambda \\ \lambda \\ f_{3x}
 \end{array}
\right),
\end{equation}
where $v_{1x}$, $v_{2x}$ and $v_{3x}$ are the velocities of the beads in response to $f_{3x}$. Using the rigid body constraint on the dimer $v_{1x}-v_{2x}=0$, 
we find
\begin{equation}
\label{eq:lambda}
 \lambda=\left \{\frac{D_\mathrm{L}^\mathrm{c}(d_{13}/\ell_0)-D_\mathrm{L}^\mathrm{c}(d_{23}/\ell_0)}{2\big[D^\mathrm{0}-D_\mathrm{L}^\mathrm{c}(d_{12}/\ell_0)\big]} \right \} f_{3x}\,.
\end{equation}
Inserting Eq.~\eqref{eq:lambda} into Eq.~\eqref{eq:dif-eq} and using 
$
 {D^\mathrm{s}_\mathrm{L}}/{(k_BT)} = {v_{3x}}/{f_{3x}}
$
yields the longitudinal self-diffusion coefficient for bead 3 in the presence of the rigid dimer.  The relative change associated with
adding the dimer as defined in Eq. \ref{eq:rel-err} is then given by
\begin{equation}
 \Delta D^\mathrm{s}_\mathrm{L}  
= \frac{\big[D_\mathrm{L}^\mathrm{c}(d_{13}/\ell_0)-D_\mathrm{L}^\mathrm{c}(d_{23}/\ell_0)\big]^2}{2D^\mathrm{0}\big[D^\mathrm{0}-D_\mathrm{L}^\mathrm{c}(d_{12}/\ell_0)\big]}\,.
\end{equation}
The above equation is plotted in Fig.~\ref{fig:dimon} for three separations.  A maximum, similar to that
appearing in Eq. \ref{fig:self-both-moving}, is clearly apparent.
%
%

\end{document}